# Biaxial strain modulated electronic structures of layered two-dimensional $MoSiGeN_4$ Rashba systems

Puxuan Li[1#], Xuan Wang[1,2#], Haoyu Wang[1], Qikun Tian[1], Jinyuan Xu[3], Linfeng Yu[3], Guangzhao Qin[3*], Zhenzhen Qin[1*]

[1]*International Laboratory for Quantum Functional Materials of Henan, and School of Physics and Microelectronics, Zhengzhou University, Zhengzhou 450001, P. R. China*
[2]*Institute for Frontiers in Astronomy and Astrophysics, Department of Astronomy, Beijing Normal University, Beijing 100875, P. R. China*
[3]*National Key Laboratory of Advanced Design and Manufacturing Technology for Vehicle, College of Mechanical and Vehicle Engineering, Hunan University, Changsha 410082, P. R. China*

# Abstract

The two-dimensional (2D) $MA_2Z_4$ family has received extensive attention in manipulating its electronic structure and achieving intriguing physical properties. However, engineering the electronic properties remains a challenge. Herein, based on first-principles calculations, we systematically investigate the effect of biaxial strains on the electronic structures of 2D Rashba $MoSiGeN_4$ (MSGN), and further explore how the interlayer interactions affect the Rashba spin splitting in such strained layered MSGNs. After applying biaxial strains, the band gap decreases monotonically with increasing tensile strains but increases when the compressive strains are applied. An indirect-direct-indirect band gap transition is induced by applying a moderate compressive strain (< 5%) in the MSGNs. Due to the symmetry breaking and moderate spin-orbit coupling (SOC), the monolayer MSGN possess an isolated Rashba spin splitting (R) near the Fermi level, which could be effectively regulated to the Lifshitz transition (L) by biaxial strain. For instance, a L←R→L transformation of Fermi surface is presented in monolayer and a more complex and changeable L←R→L→R evolution is observed in bilayer and trilayer MSGNs as the biaxial strain vary from −8% to 12%, which actually depend on the appearance, variation, and vanish of the Mexican hat band in the absence of SOC under different strains. The contribution of Mo-$d_{z^2}$ orbital hybridized with N-$p_z$ orbital in the highest valence band plays a dominant role on the band evolution under biaxial strains, where the R→L evolution corresponds to the decreased Mo-$d_{z^2}$ orbital contribution. Our study highlights the biaxial strain controllable Rashba spin splitting, in particular the introduction and even the evolution of Lifshitz transition near Fermi surface, which makes the strained MSGNs as promising candidates for future applications in spintronic devices.

# These authors contributed equally to this work.
* Corresponding author: gzqin@hnu.edu.cn; qzz@zzu.edu.cn

# I. INTRODUCTION

Two-dimensional (2D) materials represented by graphene[1,2] have attracted much attention for their novel electronic properties and distinguished performance in valleytronics[3,4], nanoelectronics[5,6], energy storage and conversion[7,8]. An increasing number of 2D materials have appeared, such as graphene and hexagonal boron nitride (*h*-BN) monolayer[9], transition metal dichalcogenides (TMDs)[10], GaSe[11] and InSe[12], multilayered $Bi_2X_3$ (X=As, Se, T)[13] and $MoSi_2N_4$ (MSN)[14]. Note that since Hong *et al.* experimentally synthesized the septuple atomic layered 2D MSN with high electron and hole mobilities and excellent ambient stability by metal-organic chemical vapor deposition (MOCVD)[14], there have been tremendous interests in MSN monolayer due to its potential application in valleytronic devices[15] and optoelectronic devices[16]. Meanwhile, the $MA_2Z_4$ family represented by the MSN are regarded as a promising 2D materials with stable structure and rich electronic characteristics[17]. As a rare class of multi-atomic layer materials, plenty of regulation ways are employed to modulate the electronic structure and physical properties of 2D MSN and other $MA_2Z_4$ systems, such as the atomic doping[18], strain engineering[19,20], and electric field modulation[20]. In addition, the construction of heterostructures or stacking layers is always be used to broaden the tunability of physical properties, including $CrCl_3$/MSN heterostructure[21], graphene/MSN heterojunction[22], MoSH/MSN heterostructures[23]. Noticed, the electrical property of $VSi_2N_4$ systems are strongly affected by stacking layers[24].

Recently, Janus 2D materials are receiving more attention for their Rashba spin splitting caused by the interplay between the intrinsic symmetry breaking and spin-orbit coupling (SOC)[10,25–29]. The use of the name "Janus" can be traced back to 1989 when C. Casagrande *et al.* prepared "Janus beads", with one hemisphere hydrophilic, and the other one hydrophobic[30]. After that, the term "Janus" has been commonly used to describe 2D materials with different atoms on each side, e.g., the typical Janus MXY (M= Mo, W; X, Y= S, Se, Te) system, where the combination of symmetry-broken structures and strong SOCs give rise to the Rashba spin splitting[10]. The Rashba effect was firstly observed in 2D electron gas by Bychkov and Rashba in 1984[31], which plays an important role in the next-generation spin-based electronics field for spin field-effect transistor[32], spin galvanic effect[33], spin hall effect[34], quantum computations[35]. The Bychkov-Rashba model, $H_R=\alpha_R(\boldsymbol{\kappa}\times\hat{\boldsymbol{z}})\cdot\boldsymbol{\sigma}$, is used to describe the Rashba effect, where $\alpha_R$ is the Rashba parameter, $\boldsymbol{\kappa}$ is the momentum, and $\hat{\boldsymbol{z}}$ is the electric field direction, $\boldsymbol{\sigma}$ is the Pauli spin matrices[31]. The rashba parameter is defined as $\alpha_R=2E_R/k_R$ , which could represent the strength of Rashba effect[36]. Since then, an increasing number of Rashba systems are discovered in other Janus materials with considerable SOC, such as Janus TMDs[10], SbTeI[28], $MoSiGeN_4$ and $WSiGeN_4$[37]. Meanwhile, various manipulating methods of the Rashba effect are also widely concerned, such as the external electric field[38], strain engineering[39,40], charge doping[41] and stacking different layers[42]. Among them, the strain engineering is widely applied to regulate the electronic structure and the Rashba effect[39,40]. For example, a large tunability (from ~0.7 eVÅ to ~1.3 eVÅ) of Rashba constant could be realized through a relatively small strain (from −2% to 2%) due to the modified orbital overlap in WSeTe monolayer[39]. Similarly the Rashba spin splitting of WSSe monolayer could be also effectively manipulated by the biaxial strain ranging from −5% to 5%[40].The 2D MSN as a representative material of $MA_2Z_4$ family, due to the coexistence advantage of rare multi-atomic layer structure, wide band gap and its exceptional stability to air/water/acid/heat, has received extensive attention in achieving intriguing physical properties and manipulating its electronic structure via the common ways of doping[37] or strain[20,43]. Noticed, Guo *et al.* designed the Janus $MoSiGeN_4$ (MSGN) and $WSiGeN_4$ systems and discovered an ideal Rashba spin splitting in valence band maximum (VBM) of MSGN[37]. Moreover, it is found that the biaxial strain is an effective way to tune the band gap and realize the indirect-direct transformation in monolayer MSN[43] and bilayer MSN[20]. Motivated by above reports naturally, it is therefore necessary and important to figure out whether the typical MSGN Rashba system is tunable under biaxial strain and stacking different layers.

In this work, we comprehensively investigate the structural stability, electronic structure and Rashba spin splitting of monolayer, bilayer and trilayer MSGNs and mainly explore the effect of biaxial strains on above systems based on the first-principles calculations. Our results suggest that biaxial strain could serve as an effective tool to regulate the band gap, isolated Rashba spin splitting, in particular the introduction and even the evolution of Lifshitz transition near Fermi surface in monolayer and multilayer MSGNs. Due to the appearance, variation and vanish of the Mexican hat band under different strains, a L←R→L transformation of Fermi surface is presented in monolayer MSGN including SOC, and a more complex and changeable L←R→L→R evolution is observed in the bilayer and trilayer MSGNs. The complex and changeable Fermi surface in such strained MSGNs is dominated by the changes in orbital contributions of Mo-$dz^2$ orbital hybridized with N-$p_z$ orbital. Our results demonstrate that layered 2D MSGNs could be promising candidates for future applications in spintronic devices.

## II. METHOD

We performed first-principles calculations by using the Vienna ab initio simulation package (VASP) based on density functional theory (DFT) computations[44]. The precise electronic properties and the ion-electron interaction were determined by using the Perdew−Burke−Ernzerhof (PBE) functional of the generalized gradient approximation (GGA)[45] and the projector-augmented wave potentials (PAWs)[44], respectively. The energy cutoff of 520 eV was applied for the geometric optimization. Structural relaxations were done with a threshold of $10^{-6}$ eV and 0.02 eV/ Å for the convergence of force and energy on each atom, respectively. The Brillouin zone was sampled by a 9×9×1 *k*-mesh for the structural relaxation, whereas for the self-consistent calculations, the Brillouin zone was performed using a dense 15×15×1 *k*-mesh. To eliminate the interaction between the two periodic MSGN systems, we used a vacuum thickness of 18 Å along the *z* direction. Additionally, the 4×4×1 supercell and 2×2×1 *k*-mesh were used for phonon spectrum calculation by suing

Phonopy code[46], and the ab initio molecular dynamics (AIMD) simulations using the NVT ensemble were carried out with the same supercell at 300 K. The van der Waals (vdW) correction was considered by adding the vdW dispersion in the DFT-D3 approximation. To calculate the spin texture, a 2D $k$-mesh ($k_X$-$k_Y$ : 60×60) was set, and PyPocar code[47] was used for the postprocessing of spin-texture data.

# III. RESULTS AND DISCUSSION

## A. Structures and stabilities of $MoSi_2N_4$ and $MoSiGeN_4$

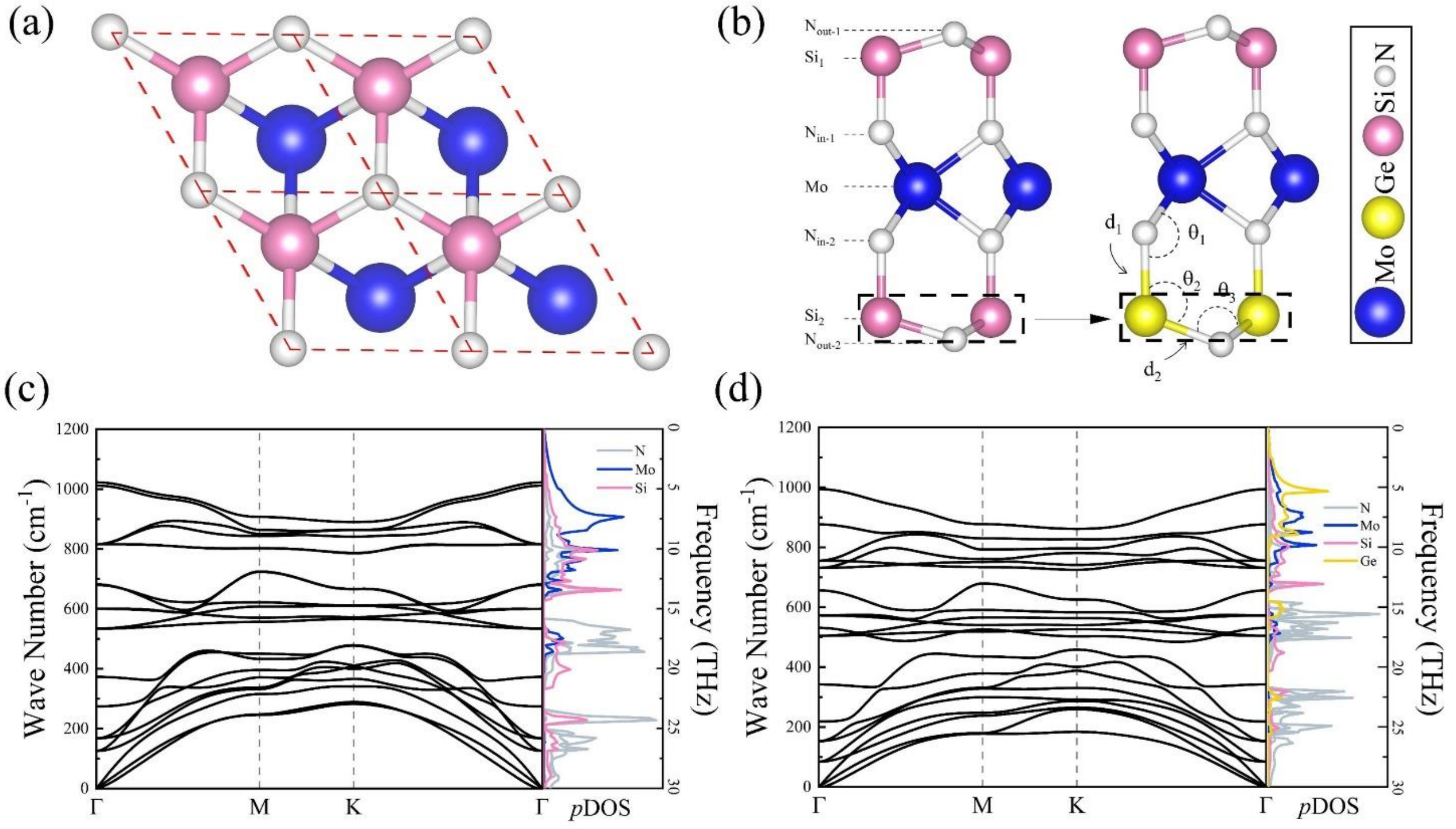

FIG. 1. Structures and stabilities of $MoSi_2N_4$ and $MoSiGeN_4$. The top view (a) and the side view (b) of monolayer $MoSi_2N_4$ and $MoSiGeN_4$. The phonon dispersion curves with the corresponding atom partial density of states ($p$DOS) of (c) $MoSi_2N_4$ and (d) $MoSiGeN_4$.

The top view and side views of the MSN crystal structures are displayed in Figs. 1(a) and 1(b), respectively, where the Mo, Si, and N atoms form a hexagonal lattice structure with a space group of $P\bar{6}m2$. The monolayer MSN comprises seven atomic layers with a N-Mo-N layer sandwiched between two Si-N layers. The optimized structural parameters of MSN, such as the lattice constant, thickness, bond lengths, and bond angles, are listed in Table I. The optimized lattice constants $a$ and thickness $t$ of MSN are 2.90 Å and 7.01 Å, respectively, which are consistent with previous calculations[16,48]. The bond lengths of Si–N ($d_1$, $d_2$) and angles of $\angle_{\text{Mo-N-Si}}$ ($\theta_1$), $\angle_{\text{N-Si-N}}$ ($\theta_2$), $\angle_{\text{Si-N-Si}}$ ($\theta_3$) are 1.74 Å, 1.75 Å, 126.64°, 106.82°, and 111.92°, respectively.

The monolayer MSGN is constructed by using the Ge layer to replace one layer of Si from the pure monolayer MSN, in which the intrinsic mirror symmetry is broken (as shown in Fig. 1(b)). After full structural relaxation, the lattice constant of MSGN, thickness, bond lengths of Ge–N ($d_1$, $d_2$) and bond angles of $\angle_{\text{Mo-N-Ge}}$ ($\theta_1$), $\angle_{\text{N-Ge-N}}$ ($\theta_2$), $\angle_{\text{Ge-N-Ge}}$ ($\theta_3$) are summarized in Table I. Noticed, the lattice constant (2.96 Å) and thickness (7.26 Å) of MSGN is relatively increased compared to that of pure MSN system, due to the increased bond length of Ge-N. Based on Bader analysis, there is more charge transfer between the Si and N atoms (2.97 e) than between the Ge and N atoms (1.87 e), leading to a more strongly polarized Si-N bond, which is consistent with the shorter Si-N bond[49].

TABLE I. Lattice constant ($a$), the layer thickness($t$), bond length($d$), angles ($\theta$), cohesive energy ($E_{\text{coh}}$) and band gap ($E_{\text{g-PBE}}$ and $E_{\text{g-PBE+vdW}}$) for $MoSi_2N_4$ and $MoSiGeN_4$. The representative bond lengths ($d_1$, $d_2$) and angles ($\theta_1$, $\theta_2$, $\theta_3$) are labeled in Fig. 1(b). Some results reported by others are also listed for comparison.

| | $a$=$b$ (Å) | $t$ (Å) | Bond (Å) | | Angle (°) | | | $E_{\text{coh}}$ (eV) | $E_g$ (eV) |
|---|---|---|---|---|---|---|---|---|---|
| | | | $d_1$ | $d_2$ | $\theta_1$ | $\theta_2$ | $\theta_3$ | | |
| $MoSi_2N_4$ | 2.91 | 7.01 | 1.74 | 1.75 | 126.64 | 106.82 | 111.92 | 6.29 | 1.77(PBE)<br>1.84(PBE+vdW)<br>3.05(HSE+vdW) |
| | 2.91[48] | 7.01[16] | 1.75[50] | 1.75[50] | -- | -- | 106[16] | 8.42[14]<br>7.34[43] | 1.74(PBE)[14]<br>2.29(HSE)[14]<br>~1.94(Exp.)[14]<br>1.74[15] |
| $MoSiGeN_4$ | 2.96 | 7.26 | 1.88 | 1.84 | 125.22 | 111.90 | 107.27 | 5.76 | 1.32(PBE)<br>1.35(PBE+vdW) |
| | 2.96[37] | -- | 1.85[50] | 1.87[50] | -- | -- | -- | 5.23[51] | 1.34[52] |

To check the energy stability of MSGN monolayer, we calculate the cohesive energy $E_{\text{coh}}$ of MSN and MSGN by the formula:

$$E_{\text{coh}} = (4E_{\text{N}} + E_{\text{Mo}} + n_1 E_{\text{Si}} + n_2 E_{\text{Ge}} - E_{\text{tot}})/7, \qquad (1)$$

where $E_{\text{tot}}$ is the total energy of monolayer MSN or MSGN; $E_{\text{Mo}}$, $E_{\text{Si}}$, $E_{\text{Ge}}$, and $E_{\text{N}}$ refer to the single atom energies of Mo, Si, Ge and N; and $n_1$, $n_2$ are the number of Si and Ge atoms in the unit cell, respectively. As listed in Table I, the $E_{\text{coh}}$ of MSGN and MSN are 5.76 eV and 6.29 eV, respectively, confirming the structural stability of the MSGN and MSN monolayers. Furthermore, the dynamic stability of MSN and MSGN monolayers is investigated by performing phonon dispersion calculations. As depicted in Figs. 1(c) and 1(d), the MSN and MSGN monolayers contain seven atoms within their primitive unit cell, resulting in three acoustic branches and eighteen optical branches in the phonon spectra. We can observe that the imaginary phonon modes are absent in the entire Brillouin zone (BZ), indicating the dynamic stability of MSN and MSGN monolayers. Besides, the calculated partial density of states ($p$DOS) reveals that in MSN and MSGN systems, the N atom dominates in the high frequency optical phonon branches. The low frequency optical phonon branches are mainly contributed by the Ge atom in MSGN and the Mo atom in MSN. Additionally, we investigated the thermal stability of MSGN using ab initio molecular dynamics (AIMD) simulations at 300 K. As illustrated in Fig. S3, the energy fluctuations are small and there is no obvious bond disruption during the entire simulation, indicating that the geometry of the MSGN monolayer is well preserved at 300K. As discussed above, the MSGN monolayer is structurally, dynamically, and thermally stable.

## B. Biaxial strain engineered electronic structures of monolayer $MoSiGeN_4$

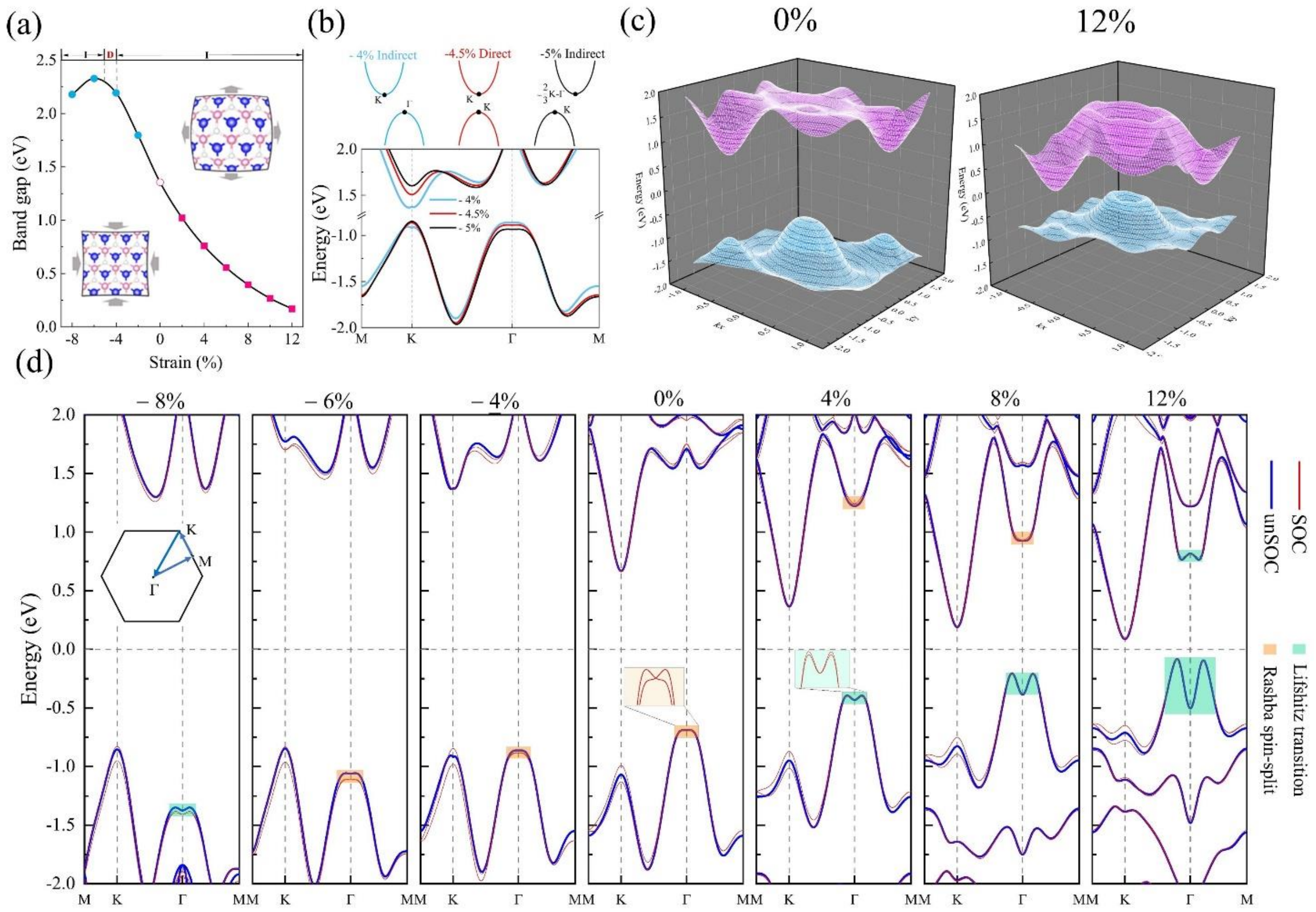

FIG. 2. Electronic structures of monolayer $MoSiGeN_4$. (a) Band gap of $MoSiGeN_4$ monolayer as a function of the biaxial strains. The simulation of compressive and tensile strains indicated as an inset. (b) The indirect (−4%)-direct (−4.5%)-indirect (−5%) band gap transition. (c) 3D plots of the monolayer $MoSiGeN_4$ band structures without and with applying the tensile strain of 12%. (d) Band structures of monolayer $MoSiGeN_4$ under biaxial strains by using PBE (blue lines) and PBE+SOC (red lines). The enlarged drawing of Rashba spin splitting (orange patch) and Lifshitz transition (green patch) are shown in the inset of (d).

The band structures of monolayer MSN are plotted using the PBE functional and the Heyd-Scuseria-Ernzerhof (HSE06) hybrid functional with considering the vdW DFT-D3 correction, as shown in Fig. S2 of the Supplemental Material. By using the PBE+vdW (HSE06+vdW) method, the parabolic conduction band minimum (CBM) and valence band maximum (VBM) locate at K point and Γ point, respectively, rendering an indirect band gap of 1.84eV (3.06 eV). It is worth mentioning that the band gap of monolayer MSN calculated by the PBE+vdW functional is closer to the experimental value(~1.94 eV[14]) without any significant change in the shape of the electronic bands compared to the HSE06+vdW functional. Therefore, the PBE functional with considering the vdW DFT-D3 correction is used for the following calculations. Similar to monolayer MSN, the monolayer MSGN is a semiconductor with an indirect band gap of 1.35 eV. Both CBM near the K point and VBM near the Γ point show parabolic nature.

In view of the adjustability of strain to the electronic structure of other 2D Janus structures[25,39,40], we further investigate the effect of biaxial strain $\varepsilon$ ($\varepsilon$ is defined as $(a-a_0)/a_0 \times 100\%$) on the geometry structures and band structures of monolayer and multilayer MSGN systems under biaxial strains changing from −8 to 12 %, where $a$ and $a_0$ denote biaxially strained and original lattice constants, with $\varepsilon<0$ ($\varepsilon>0$) being compressive (tensile) strain. As shown in Fig. S1 of the Supplemental Material, all bonds are expanded progressively under tensile biaxial strains (the horizontal bonds increase at a faster rate than the vertical bonds), and the bond angles ($\theta_1$, $\theta_2$, $\theta_5$, $\theta_7$, $\theta_8$) along the horizontal direction become larger, while the bond angles ($\theta_3$, $\theta_4$, $\theta_6$, $\theta_9$) along the vertical direction become smaller. On the other hand, all bonds are compressed gradually under compressive biaxial strains (the horizontal bonds decrease more rapidly than vertical bonds), and the horizontal bond angles decrease while the vertical bond angles increase. Therefore, the thickness of the monolayer MSGN follows a reduction trend under biaxial tensile strains and exhibits an increasing trend under biaxial compressive strains. In addition, Fig. 2(a) illustrates the evolution of band gap $E_g$ with respect to $\varepsilon$. With the increase of compressive strains, $E_g$ first increases to the maximum value (2.35 eV) when $\varepsilon = -6\%$ and then decreases monotonically, while $E_g$ decreases monotonically with increasing tensile strains. What's more, as displayed in Fig. 2(b), an indirect-direct-indirect (I→D→I) band gap transition could be induced by applying a moderate compressive strain (−4.5%). In detail, as compressive strain ranges from −4% to −4.5%, our calculations show that the CBM remains at the K point, but the VBM shifts progressively away from Γ point to K point, leading monolayer MSGN to be a direct band gap semiconductor with VBM and CBM residing at the K point. When the compressive strain further ranges from −4.5% to −5%, the position of CBM changes from K point to the two-thirds of K-Γ points, while the position of VBM remains at the K point, making the monolayer MSGN changes from the direct band gap semiconductor to the indirect band gap semiconductor. Meanwhile, the I→D→I transition is also observed at −4% with SOC. The recent work revealed that the monolayer MSN can be transformed from an indirect into a direct band gap under the strain of −4%[43]. The indirect bandgap of MSN and MSGN means that phonons need to be involved to provide momentum when electrons transition from the VBM to the CBM, which is adverse to the optoelectronic performance[53]. Therefore, from the point of view of optoelectronic device design, the efficiency of direct bandgap semiconductor devices is better than that of indirect bandgap semiconductor devices, indicating that the strained MSGN could be a promising candidate for the optoelectronic device.

To comprehensively investigate the band evolution of monolayer MSGN under biaxial strains, we calculate the band structures of monolayer MSGN under strains varying from −8% to 12% by using PBE (blue lines) and PBE+SOC (red lines) in Fig. 2(d). When the SOC is not considered, the parabolic VBM turns into the non-parabolic Mexican hat band under the tensile strains of 2% and larger compressive strains of −8% in monolayer MSGN. The 3D band structures at 0% and 12% intuitively illustrate the difference of Fermi surface brings by Mexican hat band in Fig. 2(c). Up to now, a lot of 2D materials have been predicted to have the non-parabolic Mexican hat band near Fermi surface such as $\alpha$-SnO[54], InSe[55], GaSe[56] and $In_2X_2$ (X=S, Se, Te)[57]. In general, 2D materials with Van Hove singularities induced by Mexican hat band will usually possess large thermoelectric Seebeck coefficients[58]. Moreover, both $\alpha$-SnO and GaSe exhibit tunable ferromagnetism and half-metallicity. In monolayer GaSe, the Van Hove singularity near the VBM leads to a Stoner-type magnetic instability and half-metallic ferromagnetic ground state via hole doping[56], demonstrating that the tunable magnetism is general for 2D materials with the Mexican hat band near Fermi surface. Our calculated results show that when tensile strain is applied ($\varepsilon \geq 2\%$) in monolayer MSGN, the intrinsic parabolic VBM turns into a Mexican hat band, which is consistent with the appearance of Mexican hat band in the MSN system under the tensile strain of 10%[59]. The characteristic of Mexican hat band is further intensified under the continuous increase of tensile strains to 12%. In general, the Mexican hat coefficient $M$ is defined as $M = \Delta E/\Delta K$[43], where $\Delta E$ and $\Delta K$ are the energy and momentum differences between the points of energy peak and energy valley. In the strained MSGN, $M$ increases linearly as tensile strains increase, reaching a maximum value (1.63 eV/Å) at 12%. Compared to the monolayer MSGN, the maximum value of $M$ is 0.76 eV/Å for MSN system in the same strain range[59]. Our results indicate that the monolayer MSGN doped by Ge atoms not only maintains the Mexican hat characteristics of the original MSN system under biaxial strains, but also exhibits a more remarkable modification of $M$ due to the fact that the VBM of MSN is contributed by Mo-$dz^2$ orbital[59] while that of MSGN is mainly contributed by Mo-$dz^2$ and small proportion of N-$p_z$ orbitals (as depicted in Fig. S5 of the Supplemental Material).

Inclusion of SOC, the Rashba spin splitting appears in the 2D materials with broken inversion symmetry[36,60]. As expected, in monolayer MSGN, the isolated Rashba spin splitting is found near the Γ point of VBM due to the symmetry breaking and SOC, which is consistent with Guo's report[37]. The Rashba parameter is defined as $\alpha_R = 2E_R/k_0$, where $E_R$ is the energy difference between the energies of the peak and the degenerate energy at the high symmetry point, and $k_0$ is the momentum offset between the peak and the high symmetry point[60]. We find that the $E_R$, $k_0$ and $\alpha_R$ of the MSGN monolayer is 2.15 meV, 0.07 $Å^{-1}$ and 61.1 meVÅ, respectively. Meanwhile, inclusion of SOC, the energy offset and momentum offset of Mexican hat band are seen in the highest valence band of Γ point, leading to the appearance of intraband Lifshitz transition, indicating that in the mirror-symmetry breaking 2D materials, the interaction of Rashba SOC and Mexican hat band near Fermi level couldlead to intraband Lifshitz transitions[55]. As displayed in Fig. 2(d), it is found that the Rashba spin splitting could be effectively regulated to a Lifshitz transition in monolayer MSGN. Interestingly, as biaxial strains change from −8% to 12%, a L←R→L transformation of Fermi surface is presented in monolayer MSGN.

What's more, within the moderate strain range ($-6\% \leq \varepsilon \leq 0\%$), the $\alpha_R$ of $V_1$ ($V_1$ denotes the highest valence band of the Γ point) decreases first and then increases with increasing compressive strains, until $\varepsilon$ reaches to −8%, and the Lifshitz transition is observed. When a small tensile strain is applied ($\varepsilon > 0\%$), the Rashba spin splitting of $V_1$ directly transforms into Lifshitz transition. In addition, as $\varepsilon$ vary from 2% to 8%, a new Rashba spin splitting is present in $C_1$ ($C_1$ denotes the lowest conduction band of the Γ point), and the $\alpha_R$ increases with increasing tensile strains (as listed in the Table SI of Supplemental Material).

As discussed above, in the absent of SOC, the highest valence band of Γ point in monolayer MSGN changes from parabolic band to Mexican hat band by applying biaxial strains, maintaining the Mexican hat characteristic of the pure MSN system under biaxial strains. On the other hand, when SOC is considered, the Rashba spin splitting is presented at the Γ point of VBM, while the Mexican hat band turns into Lifshitz transition. It is due to the symmetry breaking and moderate SOC of monolayer MSGN that the isolated Rashba spin splitting is observed at VBM. Therefore, the evolution from the parabolic band to the Mexican hat band under biaxial strains is the inevitable reason for the transformation from original Rashba spin splitting to Lifshitz transition in monolayer MSGN with taking SOC into consideration. Actually, the Lifshitz transition in monolayer MSGN arises from the interaction of Rashba SOC and Mexican hat band. This Lifshitz transition has also been investigated in many 2D materials, such as bilayer graphene[61,62], bilayer InSe[55], and Janus $PA_2As$ (A = Si, Ge, Sn, and Pb) monolayers[63]. However, different from the above-mentioned 2D materials, the original isolated Rashba spin splitting could be directly transformed into Lifshitz transition under a relatively small tensile strain in monolayer MSGN. Meanwhile, the change trend of the Lifshitz transition is also sensitive to different strains, especially under the range of $2\% \leqslant \varepsilon \leqslant 12\%$. Within these strain range, we quantify the variation of the Lifshitz transition in detail in Sec. Ⅲ D. The van Hove singularity induced by Mexican hat band often leads to magnetism via hole doping[56], which would benefit to the large thermoelectric Seebeck coefficient[58]. Due to the locking of the spin and momentum[64], the spin-charge conversion of 2D materials could be enhanced by intraband Lifshitz transition[55,63,65], combined with the potential application of Rashba spin splitting. Therefore, the strained monolayer MSGN with Rashba spin splitting and intranband Lifshitz transition is expected to be a strong candidate for spintronics applications.

### C. Tunable Rashba spin splitting and Lifshitz transition

To quantitatively analyze the effect of biaxial strain on the splitting strength of Rashba effect, we summarize the variation of $\alpha_R$ under a moderate strain range of $-6\% \leq \varepsilon \leq 8\%$ (as listed in Table SI). The Rashba spin splitting is observed in $V_1$ under compressive strains and $C_1$ under tensile strains. As compressive strain is applied, the $\alpha_R$ firstly decreases to 30.7 meV·Å under −2%, and then increases to 96.4 meV·Å under −6%. Different from the compressive strains, the $\alpha_R$ of $C_1$ increases monotonically within the range of $0 < \varepsilon \leq 8\%$, in which the minimum value and the maximum value are 134.3 meV·Å under 2% and 195.9 meV·Å under 8%, respectively.

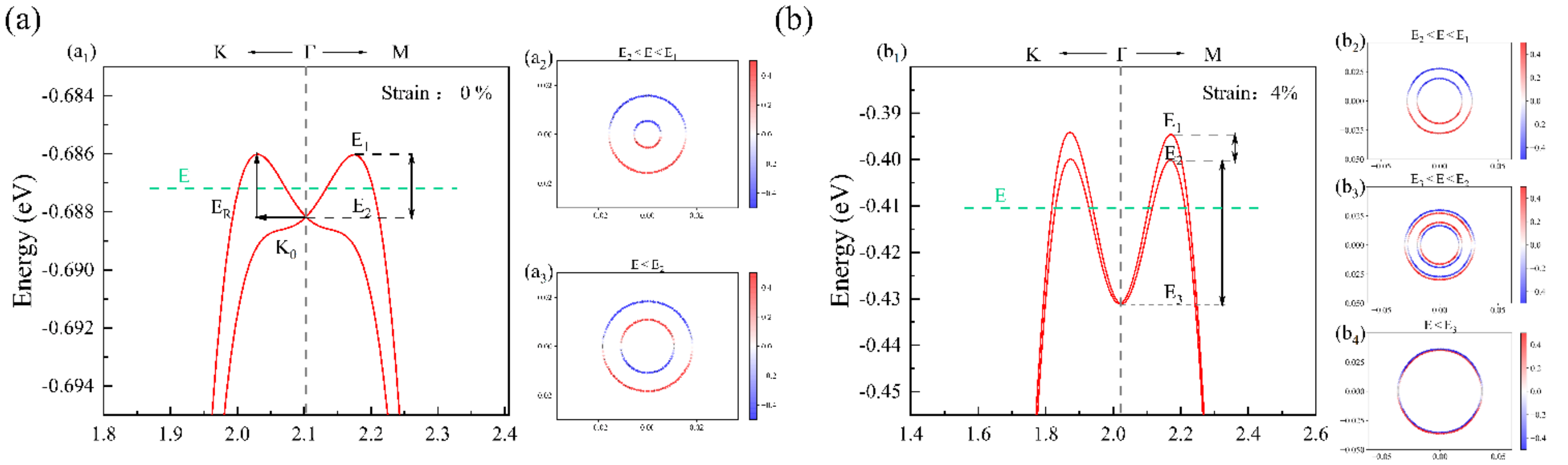

FIG. 3. Spin textures of the Rashba spin splitting and Lifshitz transition. The enlargement view of the Rashba spin splitting (a) and Lifshitz transition (b) at the VBM of $MoSiGeN_4$; Spin texture calculated in a $k_x$ - $k_y$ plane centered at the Γ point and at differents energy surfaces under 0% (a) and under 4% (b). Spin up and spin down are represented in red and blue, respectively. As for (a), $E_1$ and $E_2$ are the energy level of the peak and the valley, respectively. As for (b), $E_1$, $E_2$ and $E_3$ are the energy level of the first peak, the second peak and the valley, respectively.

The corresponding spin textures of the Rashba spin splitting and Lifshitz transition, calculated in a $k_x$ - $k_y$ plane centered near the highest valence band of Γ point under 0% and 4%, are illustrated in Figs. 3(a) and (b). As presented in Figs. 3($a_2$) and ($a_3$), $E_1$ and $E_2$ are the energy of splitting peak and Γ point, respectively. When $E_2 < E < E_1$, the spin states between the Γ-K direction and the Γ-M direction are the same. When $E < E_2$, the states with opposite spins of the Γ-K direction and the Γ-M direction appear alternately. In addition, we plot intraband Lifshitz transition spin textures within three different energy ranges (as shown in ($b_2$) ($b_3$) ($b_4$)). The energy $E_1$, $E_2$ and $E_3$ are the energy of splitting peak, the first Lifshitz transition point and the second Lifshitz transition. When $E < E_3$ and $E_2 < E < E_1$, the spin-up branch (red) and the spin-down branch (blue) form two concentric circles, which have Rashba-type spin texture. However, when $E_3 < E < E_2$, the spin texture of Lifshitz transition have four concentric circles of opposite spins states which is different from the Rashba-type spin texture. Additionally, we find that the energy difference $E_{1-2}$ and $E_{2-3}$ couldbe tuned by tensile strain, where $E_{1-2}$ is the energy range between $E_1$ and $E_2$, $E_{2-3}$ are the energy range between $E_2$ and $E_3$. Both $E_{1-2}$ and $E_{2-3}$ of monolayer MSGN linearly increase with the tensile strain varying from 2% to 12%, which will be discussed in detail in Sec. Ⅲ D.

### D. Multilayer $MoSiGeN_4$ with applying biaxial strains

Since the Si-N layer or Ge-N layer at the bottom of the upper monolayer has different interlayer interactions with the top of lower monolayer, we consider three stacking structures represented by $Si_{top}$-$Ge_{bot}$, $Ge_{top}$-$Ge_{bot}$, $Si_{top}$-$Si_{bot}$, where $Si_{top}$ denotes the Si-N layer above the van der Waals interface and $Ge_{bot}$ denotes the N-Ge layer below the van der Waals interface (as depicted in Fig. S3 of the Supplemental Material). Our calculated results demonstrate that the most preferred stacking order of the three stacking structures is $Si_{top}$-$Ge_{bot}$, which has the lowest total energy. Thus, based on the $Si_{top}$-$Ge_{bot}$ structure, we consider six different stacking patterns for bilayer MSGN inducing *AA*, *AB*, *AC*, $AA^*$, $AB^*$, $AC^*$(as shown in Fig. 4(a)). The calculated result indicates that the $AA^*$-stacking pattern has the lowest total energy. What's more, the previous work has investigated three stacking types (*AA*, *AB*, *AC*) for each bilayer structure (including $MoSiGeN_4$/$MoSiGeN_4$, $MoSiGeN_4$/$MoGeSiN_4$, and $MoGeSiN_4$/$MoSiGeN_4$), and revealed that the *AB*-stacking type based on the $MoSiGeN_4$/$MoSiGeN_4$ was the most stable structure[52]. In this work, we consider three more stacking patterns ($AA^*$, $AB^*$, $AC^*$) than Ref.[52] resulting in a more stable structure. Using the most stable structure as a reference, we work on six different stacking patterns for trilayer MSGN, i.e., *AAA*, $AAA^*$, $AA^*A$, $AA^*A^*$, $A^*AA^*$, $A^*A^*A^*$ (as shown in Fig. 4 (b)), in which the $AA^*A$-stacking pattern has the lowest total energy. Since the $AA^*$-stacking and $AA^*A$-stacking patterns exhibit the lowest total energy, we further investigate the change of the geometry structures and band structures of multilayer MSGN systems under biaxial strains changing from −8 to 12%. The AIMD of bilayer and trilayer MSGNs are plotted in Fig. S4 of the Supplemental Material within the range of $-8\% \leq \varepsilon \leq 12\%$, respectively. Our calculated results show that the bilayer and trilayer MSGNs are stable under so large strains as the total energy fluctuations are small with no structural disruption and bond breaking.

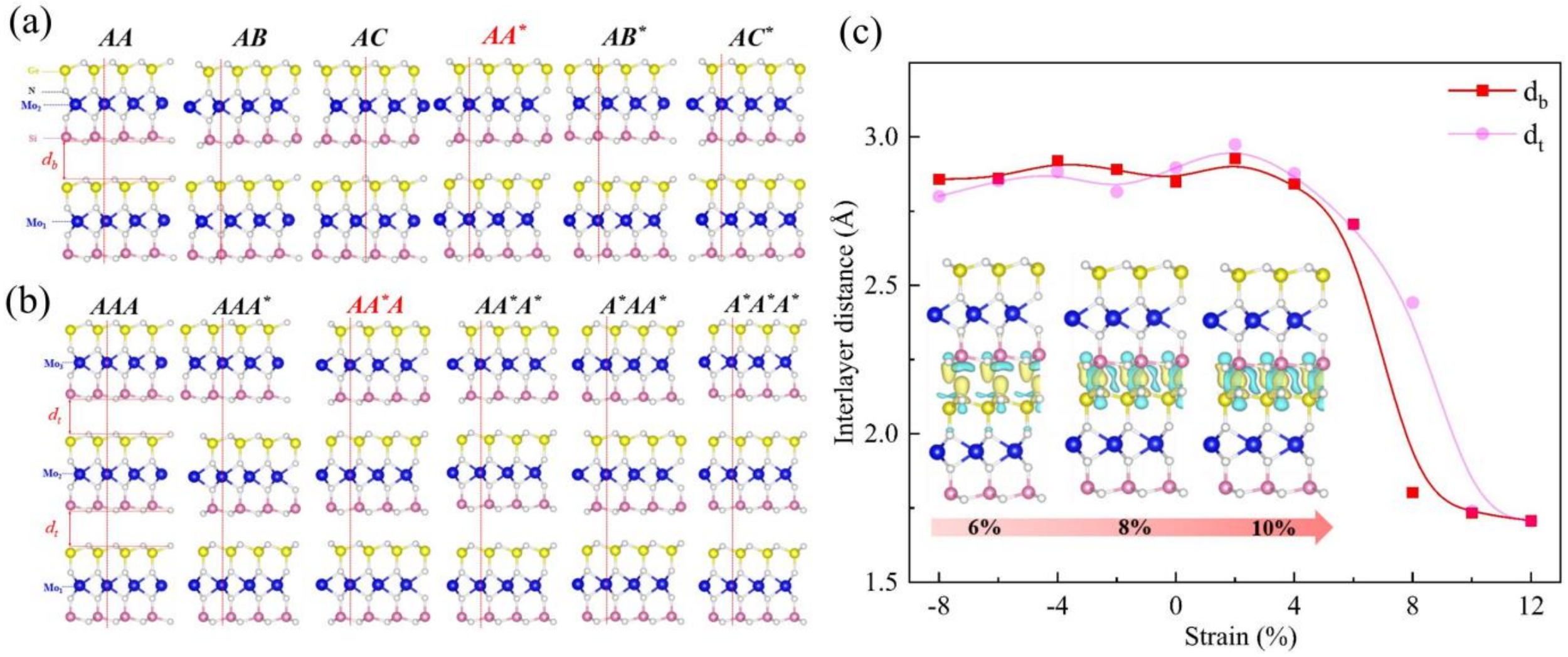

FIG. 4. Structural properties of bilayer and trilayer $MoSiGeN_4$. (a)The six different crystal structures of bilayer $MoSiGeN_4$ with *AA*, *AB*, *AC*, *AA**, *AB**, *AC** stacking modes, and (b)trilayer $MoSiGeN_4$ with *AAA*, *AAA**, *AA*A*, *A A*A**, *A*AA**, *A* A* A** stacking modes. The Mo atoms from the lowest to the highest layers are labeled as $Mo_1$, $Mo_2$, and $Mo_3$. (c)The interlayer distance of bilayer $d_b$ (red line and symbol) and trilayer $d_t$ (pink line and symbol). The structures and distributions of charge density are shown in the inset of (c), and the yellow and blue colors represent the accumulation and depletion of electrons, respectively.

Figs. 4(a) and (b) indicate that the interlayer distances $d_b$ (the vertical distance between the two N atoms of the van der Waals interface) and $d_t$ are both 2.90 Å for bilayer MSGN and trilayer MSGN, indicating that the interlayer distances remain constant even if the number of layers increases. However, unusual variation of interlayer distance appears when strain ranges from −8% to 12%. As shown in Fig. 4(c), $d_b$ and $d_t$ decrease slowly under compressive strain while very sensitive to tensile strain in multilayer MSGNs. As tensile strain is applied, the interlayer distance first increases slowly and then decreases to 2.70 Å under 6%. Further increasing strain from 6% to 8%, the $d_b$ decreases rapidly from 2.70 Å to 1.80 Å. The $d_t$ decrease drastically from 2.70, 2.44 Å to 1.74 Å under the biaxial strain of 6%, 8% and 10%, respectively.

As discussed above, the significant tunability of interlayer distance is achieved through a relatively small strain, which could be attributed to charge redistribution resulting from the biaxial strains. In detail, the Ge-N bond is 0.15 Å longer than the Si-N bond, allowing the N atoms connected to the Ge atom in the lower layer to easily interact with the Si atoms in the upper layer, thereby altering the charge distribution and causing the charge center of the N atom to move upward. Additionally, the N atoms in the upper layer could readily interact with the Ge atoms in the lower layer, causing the charge center of the N atoms to move downward. As a result, the remarkable reduction of interlayer distance is due to the charge redistribution induced by biaxial strain. The previous work also revealed that the interlayer distance of bilayer 2H-$MoS_2$ remains almost constant except for a sharp increase under the compressive biaxial strain of −6%[66]. Besides, the interlayer distances of $MoA_2N_4$ (A=Si, Ge) is 2.81Å, which is much smaller than bilayer $VSi_2N_4$ (2.98 Å)[24] due to the stronger interlayer interactions. Therefore, the sharp reduction of interlayer distances in bilayer and trilayer MSGNs may be attributed to the stronger interlayer interactions due to charge redistribution.

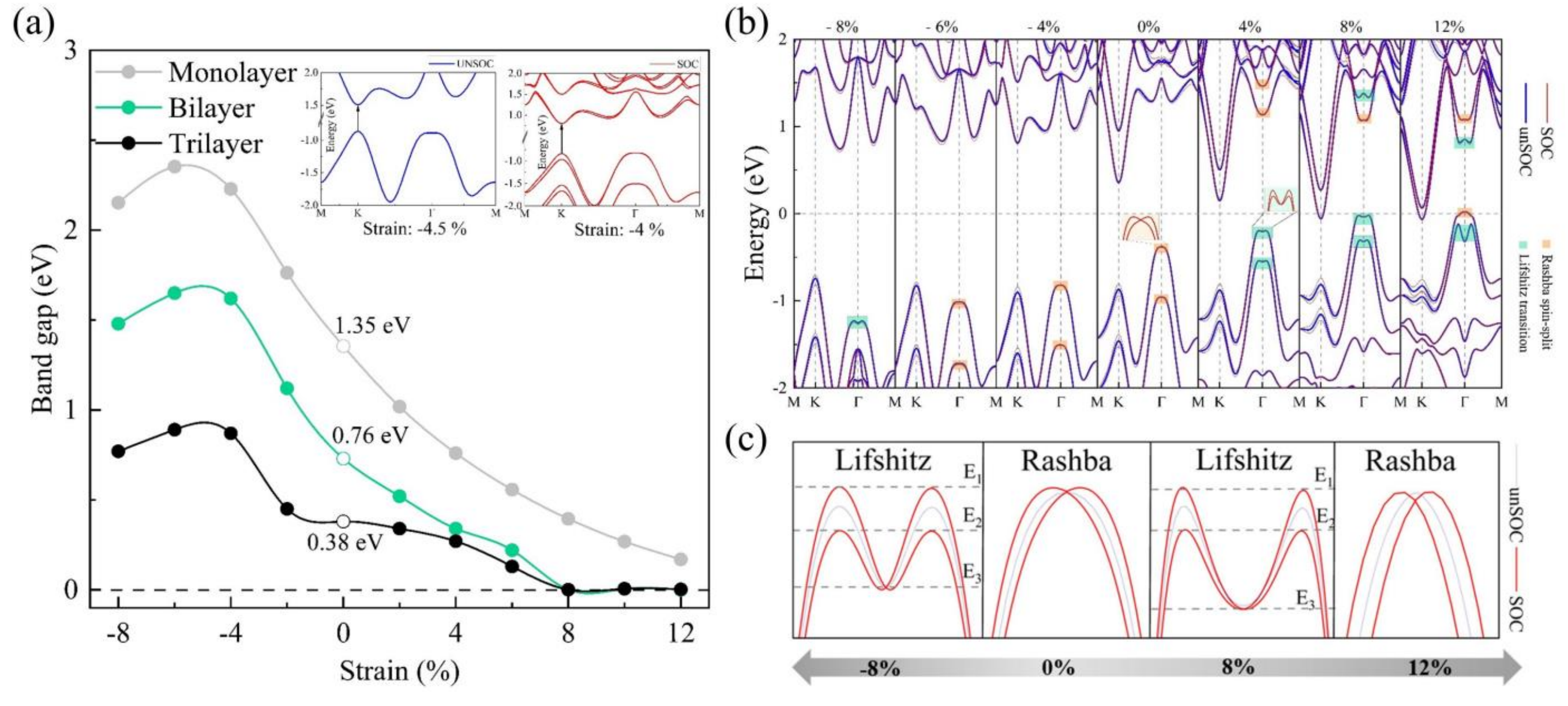

FIG. 5. Electronic properties of bilayer $MoSiGeN_4$. (a) The change of bang gap when the biaxial strain varies from −8% to 12% in monolayer (dark line and symbol), bilayer (green line and symbol), and trilayer (black line and symbol) $MoSiGeN_4$. (b) Band structure of bilayer $MoSiGeN_4$ under biaxial strains considering without SOC (blue lines) and with SOC (red lines). The enlarged drawing of Rashba spin splitting (orange patch) and Lifshitz transition (green patch) are shown in the inset of (b). (c) The schematic of Lifshitz←Rashba→Lifshitz→Rashba evolution of multilayer $MoSiGeN_4$.

Similar to the monolayer MSGN, both bilayer MSGN and trilayer MSGN are indirect bandgap semiconductors with CBM at K the point and VBM at the Γ point. As exhibited in Fig. 5(a), the bilayer MSGN with an indirect band gap of 0.73 eV is smaller than the band gap of monolayer MSGN (1.35 eV) and larger than the band gap of trilayer MSGN (0.38eV). Our calculated results show that the band gaps of MSGN decrease monotonically with an increasing number of layers, which is in agreement with the MSN multilayer systems[14], indicating that the Janus phase retains the same regulation with respect to pristine MSN systems. This trend is related to the orbital contributions of $Mo_1$ layer, $Mo_2$ layer in bilayer MSGN and $Mo_1$ layer, $Mo_2$ layer and $Mo_3$ layer in trilayer MSGN. As shown in Fig. S6 of the Supplemental Material, if the VBM of the first layer is closer to the Fermi energy level, then the CBM must be further away from the Fermi energy level. For instance, in bilayer MSGN, the VBM is contributed by $Mo_1$-$dz^2$ of the lower layer, resulting in the CBM of the lower layer being far away from the Fermi energy level, and the CBM is contributed by $Mo_2$-$dz^2$ of the upper layer, resulting in the VBM of the upper layer being far away from the Fermi energy level. As a result, the VBM of bilayer MSGN is contributed by $Mo_1$-$dz^2$ of the lower layer, and the CBM is contributed by $Mo_2$-$dz^2$ of the upper layer. And for trilayer MSGN, the VBM the is contributed by $Mo_1$-$dz^2$ of the lowest layer, and the CBM is contributed by $Mo_3$-$dz^2$ of the uppermost layer. As discussed above, the VBM and CBM of multilayer MSGNs are contributed by Mo-$dz^2$ of different layers, which is the reason for the decrease of band gaps from monolayer to trilayer.

In addition, the change of bang gap $E_g$ when the biaxial strain varies from −8% to 12% in MSGN systems is depicted in Fig. 5(a). Under tensile strain, the $E_g$ of multilayer MSGNs decrease monotonically and become 0 eV when the strain reaches 10%, transforming the bilayer MSGN and trilayer MSGN from the semiconductor into metal. Under compressive strain, the $E_g$ of bilayer MSGN and trilayer MSGN increase to the maximum value when $\varepsilon = -6\%$ and then decreases monotonically. Note that both bilayer MSGN and trilayer MSGN exhibit an I→D→I transition at −4.5% with PBE and −4% with PBE+SOC, which is consistent with monolayer MSGN. Our calculated results also indicate that $E_g$ shows similar trends in layered MSGNs under biaxial strains, where the $E_g$ first increases and then decreases monotonically under compressive strain, while continuous decreases with the tensile strain increases. Besides, an I→D→I band gap transition could be induced by applying a moderate compressive strain for all layered systems. Fig. 5(b) presents the band structure of bilayer MSGN by using PBE (blue lines) and PBE+SOC (red lines) within the range of $-8\% \leq \varepsilon \leq 12\%$. It is shown that the parabolic valence band near Fermi surface would turn to the nonparabolic Mexican hat by applying certain tensile strain (2%) or compressive strain (−8%) without considering SOC. When the tensile strain further increases to 12%, the Mexican hat changes back to the parabolic VBM. When the SOC is included, the Rashba spin splitting is revealed at the highest valence band of the Γ point under $\varepsilon$ changing from −6% to 0%. As the compressive strain increased, the Rashba spin splitting turns into Lifshitz transition. As tensile strain is applied, the intrinsic Rashba spin splitting turns into Lifshitz transition, until $\varepsilon = 12\%$, and the Rashba spin splitting appears again. Meanwhile, a new Rashba spin splitting is revealed in the lowest conduction band of the Γ point under the range of $2\% \leq \varepsilon \leq 8\%$, and $\alpha_R$ increases with increasing tensile strains (as listed in the Table SI of Supplemental Material). Therefore, compared to the L←R→L transformation of monolayer MSGN, a more complex and variable L←R→L→R evolution is observed with SOC in the highest valence band of the Γ point of bilayer MSGN. What's more, the isolated Rashba spin splitting is observed at the Γ point of VBM in trilayer MSGN. As the number of layered MSGN increases from monolayer to trilayer, $\alpha_R$ decreases from 61.1 meV·Å, 37.4 meV·Å to 19.9 meV·Å, respectively. When biaxial strain changes from −8% to 12%, an L←R→L→R evolution is also observed near the Fermi surface in trilayer MSGN, as shown in Fig. S7 of the Supplemental Material. Under tensile strains, the R→L transformation is presented in the monolayer MSGN, as well as the R→L transformation in bilayer and trilayer MSGNs, and even the reverse transformation of L→R is observed under the strain of 12%, depend on the appearance, variation and vanish of the Mexican hat band under different strains. In addition, we also summarize the calculated band structures by using PBE functional without considering the vdW DFT-D3 correction, as shown in Fig. S8 of the Supplemental Material. When the vdW interactions in MSGN system are not considered, the energy band structure of the monolayer remains constant, while a new band structure, multiple Lifshitz transition ($L_m$), appears under 10% in the bilayer MSGN, leading to a more complex L←R→L→$L_m$→R evolution in the highest valence band of the Γ point (as shown in Fig. S9).

Moreover, the atomic orbital projected band structures are calculated within the range of $-8\% \leq \varepsilon \leq 12\%$ for monolayer and bilayer MSGNs. When SOC is not included (as depicted in Fig. S5 of the Supplemental Material), we notice that the VBM in monolayer MSNG is mainly contributed by the out-of-plane Mo-$dz^2$ and small proportion of N-$p_z$ orbitals. The highest valence band of K point is contributed by in-plane N-$px,y$ orbitals. As to be expected in Fig 6(a), the Rashba spin splitting and Zeeman effect are found at VBM and the highest valence band of K point when SOC is considered. Meanwhile, we focus on the orbital contributions of the highest valence band of Γ point under biaxial strain ranging from −8% to 12%.

As displayed in Fig. 6(a), under the strain range of $-4\% < \varepsilon \leq 0\%$, the highest valence band of Γ point of the monolayer MSNG is mainly contributed by the Mo-$dz^2$ and small proportion of N-$p_z$ orbitals, where exhibits Rashba spin splitting. When compressive strain changes from −4% to −8%, the VBM changed from the Γ point to the K point of the highest valence band. In this strain range, the VBM of the monolayer MSNG is contributed by the N-$px,y$ orbitals, where exhibits Zeeman effect. What's more, the highest valence band of Γ point exhibits a Lifshitz transition under $\varepsilon = -8\%$, which is mainly contributed by out-of-plane Mo-$dz^2$ and small proportion of N-$p_z$ and in-plane N-$p_{xy}$ orbitals. In addition, the contribution of Mo-$dz^2$ and N-$p_z$ orbits of monolayer MSGN decreases with enhanced tensile strains and compressive strains, leading to an increase in $E_{1\text{-}2}$ and $E_{2\text{-}3}$ energy intervals. By further increasing the strain to 12% in monolayer MSGN, the Mo-$dz^2$ and N-$p_z$ orbits contribute the least at VBM, where the $E_{1\text{-}2}$ and $E_{2\text{-}3}$ reach the maximum value. As for bilayer MSGN (displayed in Fig. 6(b)), there are Rashba spin splitting in the highest valence band of Γ point within the range $-6\% \leq \varepsilon \leq 0\%$, which is mainly contributed by out-of-plane Mo-$dz^2$ and small proportion of N-$p_z$ orbitals. When the proportion of Mo-$dz^2$ decrease, the Lifshitz transition appears under $\varepsilon = -8\%$. When the tensile strain is applied, the Rashba spin splitting turns into Lifshitz transition due to the decreased contribution of Mo-$dz^2$. Notably, after the tensile strain reaches 12%, the Rashba spin splitting reappears, which is mainly contributed by the increased contribution of Mo-$dz^2$ orbital and N-$p_z$ orbital. For the monolayer MSGN, the proportion of Mo-$dz^2$ orbital decreases with enhanced tensile strains and compressive strains. Meanwhile, the proportion of N-$p_z$ orbital increase with a compressive strain and decrease with increasing tensile strain. Within the range of $-8\% \leq \varepsilon \leq 6\%$, the variation trend of the Mo-$dz^2$ and N-$p_z$ orbitals are consistent with the change rule of monolayer MSGN. When $\varepsilon \geq 6\%$, the decreasing trend of Mo-$dz^2$ and N-$p_z$ orbitals becomes increasing due to the reduction of interlayer distances in the bilayer MSGN.

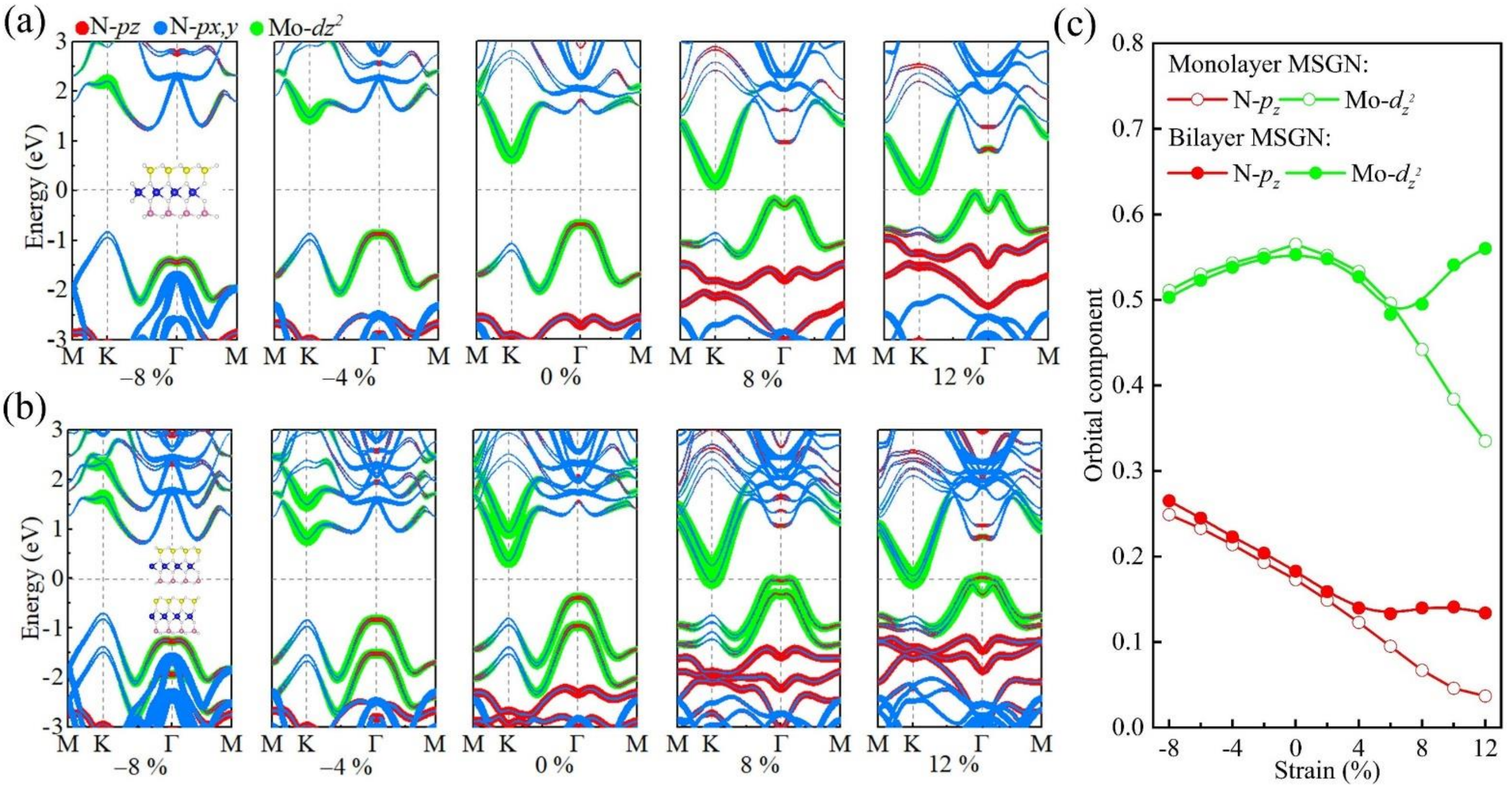

FIG. 6. Orbital-projected band structures of monolayer and bilayer $MoSiGeN_4$. (a)The orbital-projected band structures of monolayer $MoSiGeN_4$ under biaxial strains and (b) bilayer $MoSiGeN_4$ with SOC by using PBE, and the size of solid circle is proportional to the orbital weight. The color of red is the contribution of N-$p_z$. The color of blue is the contribution of N-$p_{x,y}$. The color of green is the contribution of Mo-$dz^2$. (c) The changes of N-$p_z$ and Mo-$dz^2$orbital compositions in the highest valence band of Γ point of monolayer and bilayer $MoSiGeN_4$ as functions of the biaxial strains.

As discussed above, the magnitude of the Mo-$dz^2$ orbital contribution plays a dominant role in the highest valence band of Γ point under biaxial strains. In the highest valence band of Γ point of the monolayer MSGN, the contribution of Mo-$dz^2$ orbital hybridized with N-$p_z$ orbital in the highest valence band plays a dominant role on the band evolution under biaxial strains, where the R→L evolution corresponds to a decreased Mo-$dz^2$ orbital contribution. The intrinsic Rashba spin splitting corresponds to most contribution of Mo-$dz^2$ orbital, and Lifshitz transition corresponds to less contribution of Mo-$dz^2$ orbital. The band evolution of the bilayer MSNG further validates our discussion with the band evolution of Rashba to Lifshitz transition under compressive strains, the Rashba to Lifshitz transition under tensile strains, and even the Rashba spin splitting reappeared under $\varepsilon$ = 12%, exactly corresponding to the changes of Mo-$dz^2$ orbital contribution under biaxial strains (as displayed in Fig. 6(c)).

As described Mexican hat band in Sec. III B, multilayer MSGNs also exhibits Mexican hat bands near the highest valence band of Γ point in the absence of SOC. As depicted in Fig. 7(b), when SOC is considered, the Lifshitz transition is observed in monolayer and multilayer MSGNs. In order to quantitatively describe the

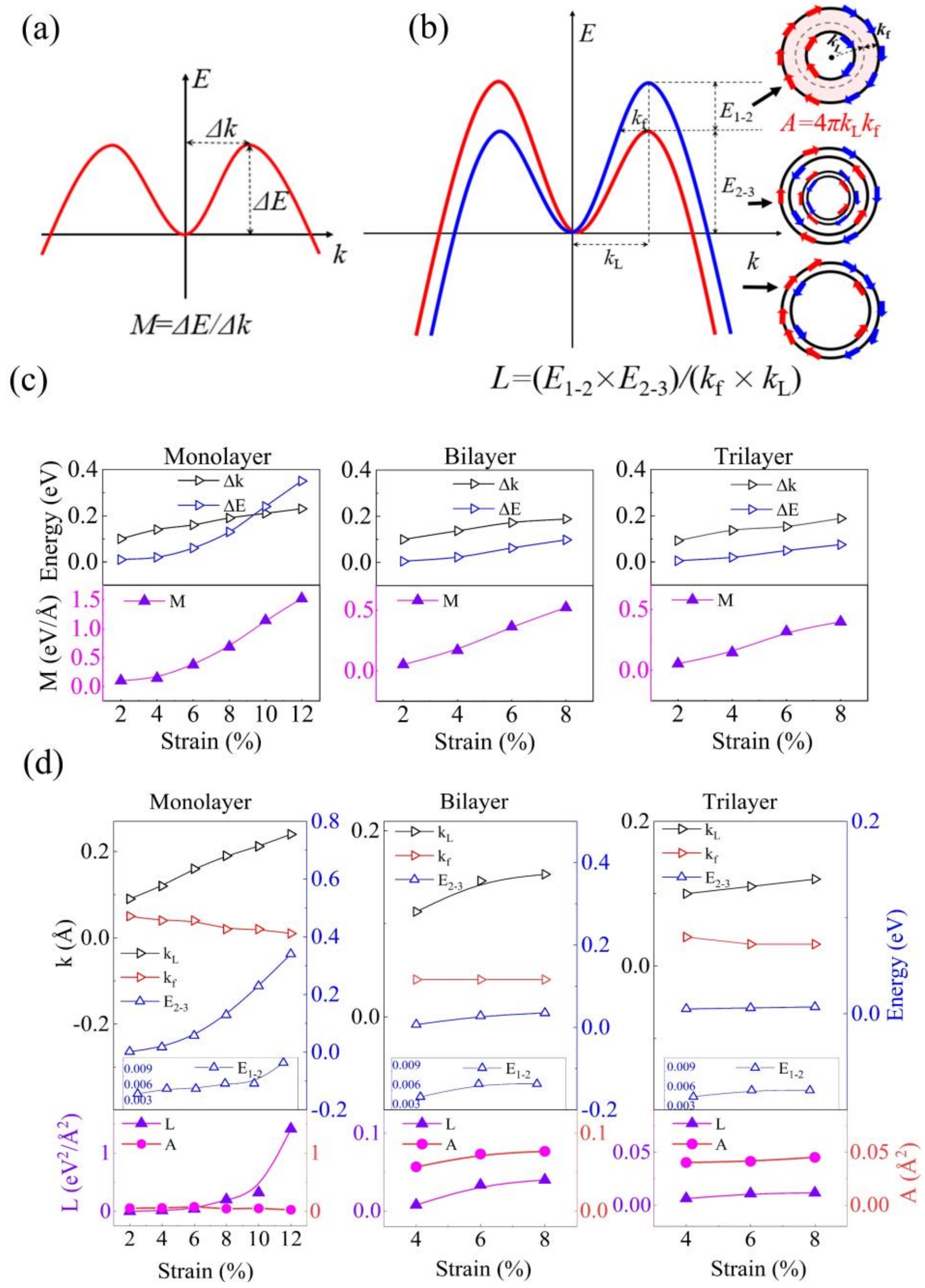

FIG. 7. Mexican-hat band and Lifshitz transition coefficients. (a) The schematic of Mexican-hat band structure. (b) The schematic of Lifshitz transition band structure. $E_{1\text{-}2}$ is the energy difference between the first and the second energy peak. $E_{2\text{-}3}$ is the energy difference between the second energy peak and energy valley. $k_f$ and $k_L$ are the momentum difference between the energy peak and energy valley points. $A$ corresponds to the area of the red circular. The spin texture is shown when energy locates in the different energy rang. (c) The Mexican-hat coefficient $M$, $\Delta k$ and $\Delta E$ of $MoSiGeN_4$ monolayer, bilayer and trilayer as a function of the strain. (d) The Lifshitz transition coefficient $L$, $E_{1\text{-}2}$, $E_{2\text{-}3}$, $k_f$, $k_L$ and $A$ of $MoSiGeN_4$ monolayer, bilayer and trilayer as a function of the strain.

continuous change of Lifshitz transition under tensile strain, a fundamental expression is introduced:

$$L = (E_{1\text{-}2} \times E_{2\text{-}3}) / (k_f \times k_L), \quad (2)$$

where $E_1$ is the energy peak, $E_2$ is the first Lifshitz transition point, $E_3$ is the second Lifshitz transition point, $E_{1\text{-}2}$ and $E_{2\text{-}3}$ are the energy range, $k_f$ and $k_L$ are the momentum range. As illustrated in Fig. 7(c), the schematic of spin texture consists of two concentric circles in the energy interval of $E_{1\text{-}2}$, in which $K_f$ and $K_L$ could be regarded as the maximum radius and radius difference of the concentric circles, respectively. We summarize the variations of $L$, $E_{1\text{-}2}$, $E_{2\text{-}3}$, $k_L$, and $k_f$ within the range of $2\% \leq \varepsilon \leq 12\%$ in monolayer MSGN, and the range of $4\% \leq \varepsilon \leq 8\%$ in multilayer MSGNs (as displayed in Fig. 7(d)). In monolayer MSGN, $E_{1\text{-}2}$ increases slowly with increasing the tensile strain, but $E_{2\text{-}3}$ and $L$ has a faster growth under tensile strain. The $k_L$ rising with an increasing tensile strain, while $k_f$ falling with an increasing tensile strain. In multilayer MSGNs, the $L$, $E_{1\text{-}2}$, $E_{2\text{-}3}$ and $k_L$ increases monotonically while $k_f$ decreases monotonically under tensile strains. Based on the introduced expression of $L = (E_{1\text{-}2} \times E_{2\text{-}3})/(k_f \times k_L)$, the variation trend of intraband Lifshitz transition could be given a more complete overview to a certain extent, and thus we could quantitatively describe the variation trend of Lifshitz transition under continuous strains. Therefore, our results will help subsequently to provide criteria for quantifying the evolution of Lifshitz transition for the analysis of other systems in which the Lifshitz transition is presented through appropriate external modulation. In addition, we calculated the value of $A$ by using $A = 4\pi k_L k_f$, corresponding to the area of the red circular region plotted in Fig. 7(c). According to the previous studies[55,63,65], $A$ could co-quantified a correlation between the variation trend of Lifshitz transition with the spin-charge conversion to a certain extent. Under tensile strains, the $A$ of monolayer, bilayer and trilayer MSGNs vary from 0.03 $Å^2$ to 0.08 $Å^2$, from 0.05 $Å^2$ to 0.07 $Å^2$ and from 0.04 $Å^2$ to 0.05 $Å^2$, respectively. Therefore, our calculated results indicate that the change of fermi surface in strained MSGN systems might be beneficial to the spin-charge conversion[55,63,65].

# CONCLUSION

In conclusion, the structural stability and electronic structure of monolayer and multilayer MSGNs have been comprehensively investigated under biaxial strains using first-principles calculations. The calculated $E_{coh}$, phonon dispersion curves and AIMD simulations reveal that the MSGN monolayer is structurally, dynamically, and thermally stable. We found the $E_g$ shows similar trends in monolayer and multilayer MSGNs under biaxial strains, where the $E_g$ first increases to the maximum under $\varepsilon = -6\%$ and then decreases monotonically under compressive strain, while continuously decreases as the tensile strain increases. Besides, an I→D→I transition is observed under a moderate strain of −4.5% without SOC and −4% with SOC in monolayer and multilayer MSGNs. As for monolayer MSGN, the highest valence band of monolayer MSGN changes from parabolic band to Mexican hat band under a small strain of 2% and a moderate compressive strain of −8%, maintaining the Mexican hat characteristic of the pure MSN system under biaxial strains. When SOC is considered, the parabolic band turns into Rashba spin splitting at the Γ point of VBM, and the Mexican hat band turns into Lifshitz transition. It is noticed that, the Rashba spin splitting could be effectively regulated to a Lifshitz transition under biaxial strains in monolayer MSGN, in which a L←R→L transformation of the Fermi surface is presented. For bilayer and trilayer MSGNs, a more complex and changeable L←R→L→R evolution is observed under the biaxial strain varying from −8% to 12%, which actually depend on the appearance, variation, and vanish of the Mexican hat band in the absence of SOC under different strains. Analyzed from the orbital-projected band structures, we found that the complex and changeable Fermi surface in such strained MSGNs is dominated by the changes in orbital contributions of Mo-$dz^2$ orbital hybridized with N-$p_z$ orbital, where the R→L transformation corresponds to a decreased contribution of Mo-$dz^2$ orbital, and L→R transformation corresponds to an increased contribution of Mo-$dz^2$ orbital. Additionally, we introduce a primary expression $L = (E_{1\text{-}2} \times E_{2\text{-}3})/(k_f \times k_L)$ in order to quantitatively describe the continuous change of Lifshitz transition under tensile strains, and further calculate $A = 4\pi k_L k_f$ of strained monolayer and multilayer MSGNs which could co-quantify a correlation between the variation trend of Lifshitz transition with the spin-charge conversion to a certain extent. Therefore, the strained monolayer and multilayer MSGN systems with tunable Rashba spin splitting and Lifshitz transition are expected to be strong candidates for spintronics applications.

## Acknowledgements

Z. Q. is supported by the National Natural Science Foundation of China (Grant No. 12274374, 11904324) and the China Postdoctoral Science Foundation (2018M642774). G. Q. is supported by the National Natural Science Foundation of China (Grant No. 52006057), the Fundamental Research Funds for the Central Universities (Grant No. 531119200237 and 541109010001). The numerical calculations in this work are supported by National Supercomputing Center in Zhengzhou.

*Supplemental Material*

# Biaxial strain modulated electronic structure of layered two-dimensional $MoSiGeN_4$ Rashba systems

Puxuan Li[1#], Xuan Wang[1,2#], Haoyu Wang[1], Qikun Tian[1], Jinyuan Xu[3], Linfeng Yu[3], Guangzhao Qin[3*], Zhenzhen Qin[1*]

[1]*International Laboratory for Quantum Functional Materials of Henan, and School of Physics and Microelectronics, Zhengzhou University, Zhengzhou 450001, P. R. China*

[2]*Institute for Frontiers in Astronomy and Astrophysics, Department of Astronomy, Beijing Normal University, Beijing 100875, P. R. China*

[3]*National Key Laboratory of Advanced Design and Manufacturing Technology for Vehicle, College of Mechanical and Vehicle Engineering, Hunan University, Changsha 410082, P. R. China*

[#] These authors contributed equally to this work.

[*] Corresponding author: gzqin@hnu.edu.cn; qzz@zzu.edu.cn

## (a) The change of Geometry Structures under biaxial strain

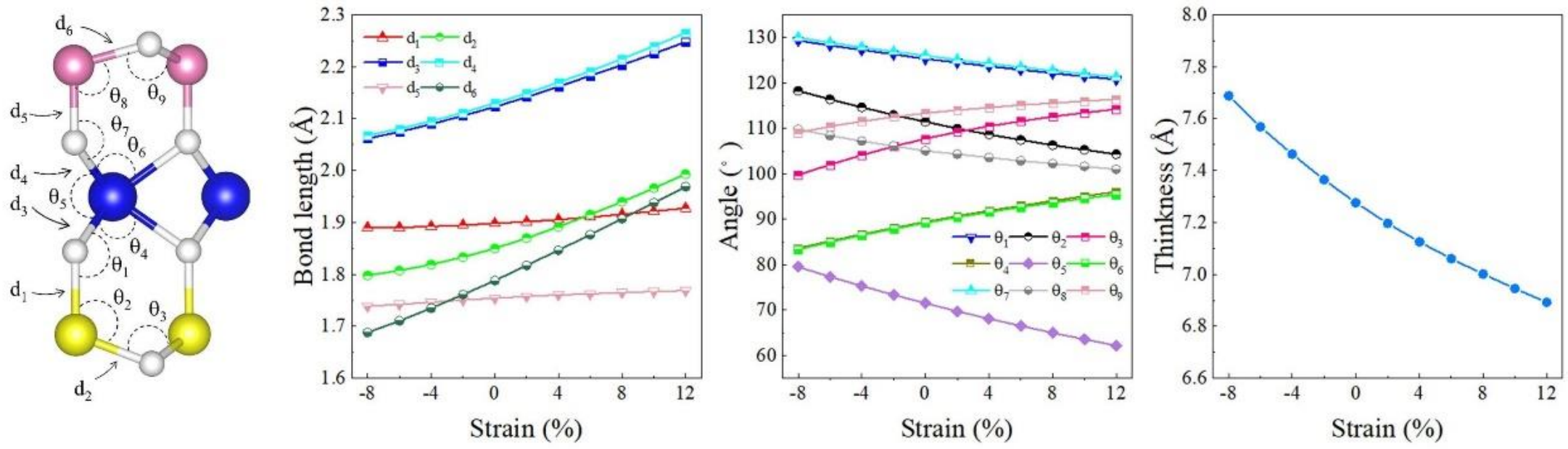

Fig. S1. The schematic diagram of structural parameters and the changes of band length, band angle and thickness under biaxial strains. The thickness and $\theta_1$, $\theta_2$, $\theta_5$, $\theta_7$, $\theta_8$ of $MoSiGeN_4$ increases with the decrease of compressive strain and decreases as tensile strain increase, a trend opposite to that of bonds length and $\theta_3$, $\theta_4$, $\theta_6$, $\theta_9$.

## (b) Band structure of $MoSi_2N_4$ and $MoSiGeN_4$ by using HSE and PBE

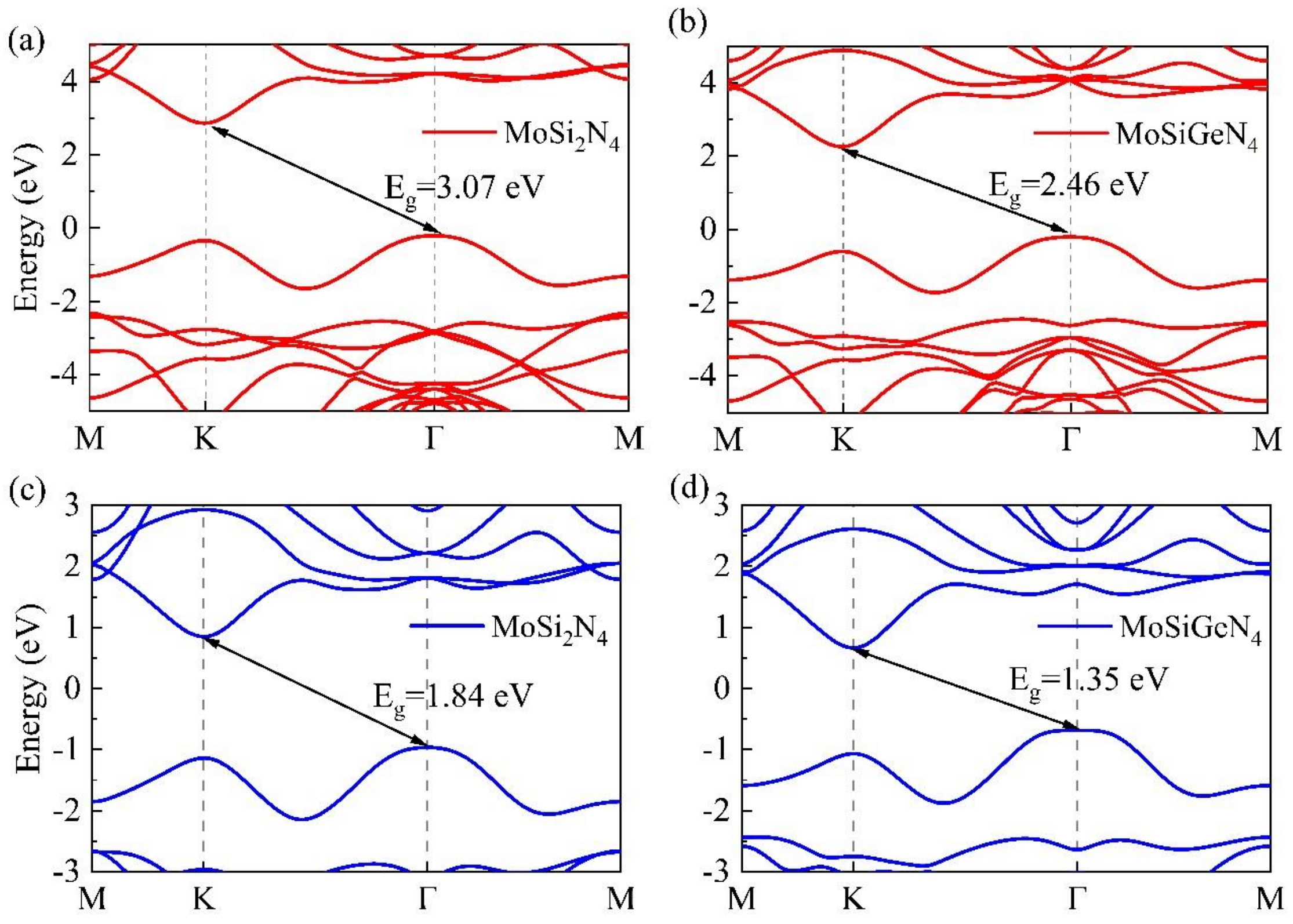

Fig. S2. The band structure of $MoSi_2N_4$ (a) and $MoSiGeN_4$ (b) by using HSE functional (red line). The band structure of $MoSi_2N_4$ (c) and $MoSiGeN_4$ (d)by using PBE functional (blue line).

## (c) The Geometry Structures of bilayer $MoSiGeN_4$

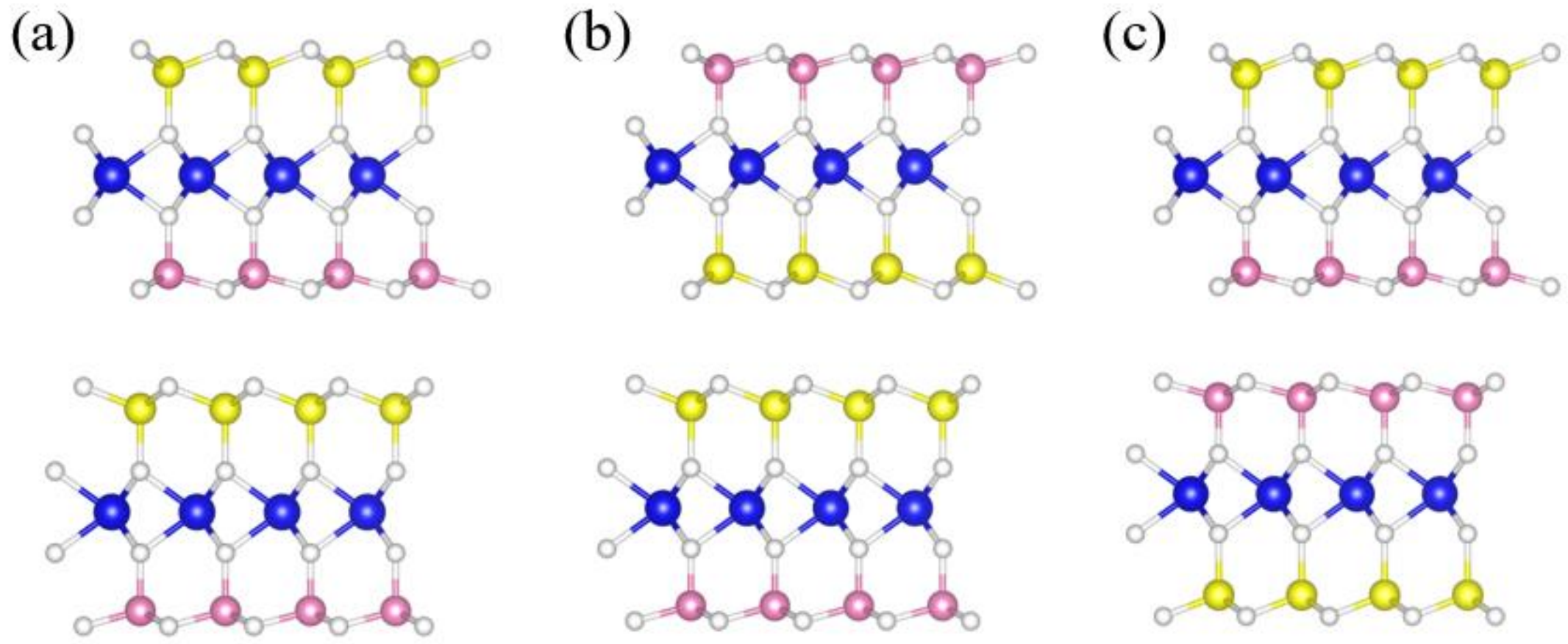

Fig. S3. The three stacking structures of (a) $Si_{top}$-$Ge_{bot}$ (b) $Ge_{top}$-$Ge_{bot}$ and (c) $Si_{top}$-$Si_{bot}$ for bilayer $MoSiGeN_4$.

## (d)Ab-initio Molecular Dynamics

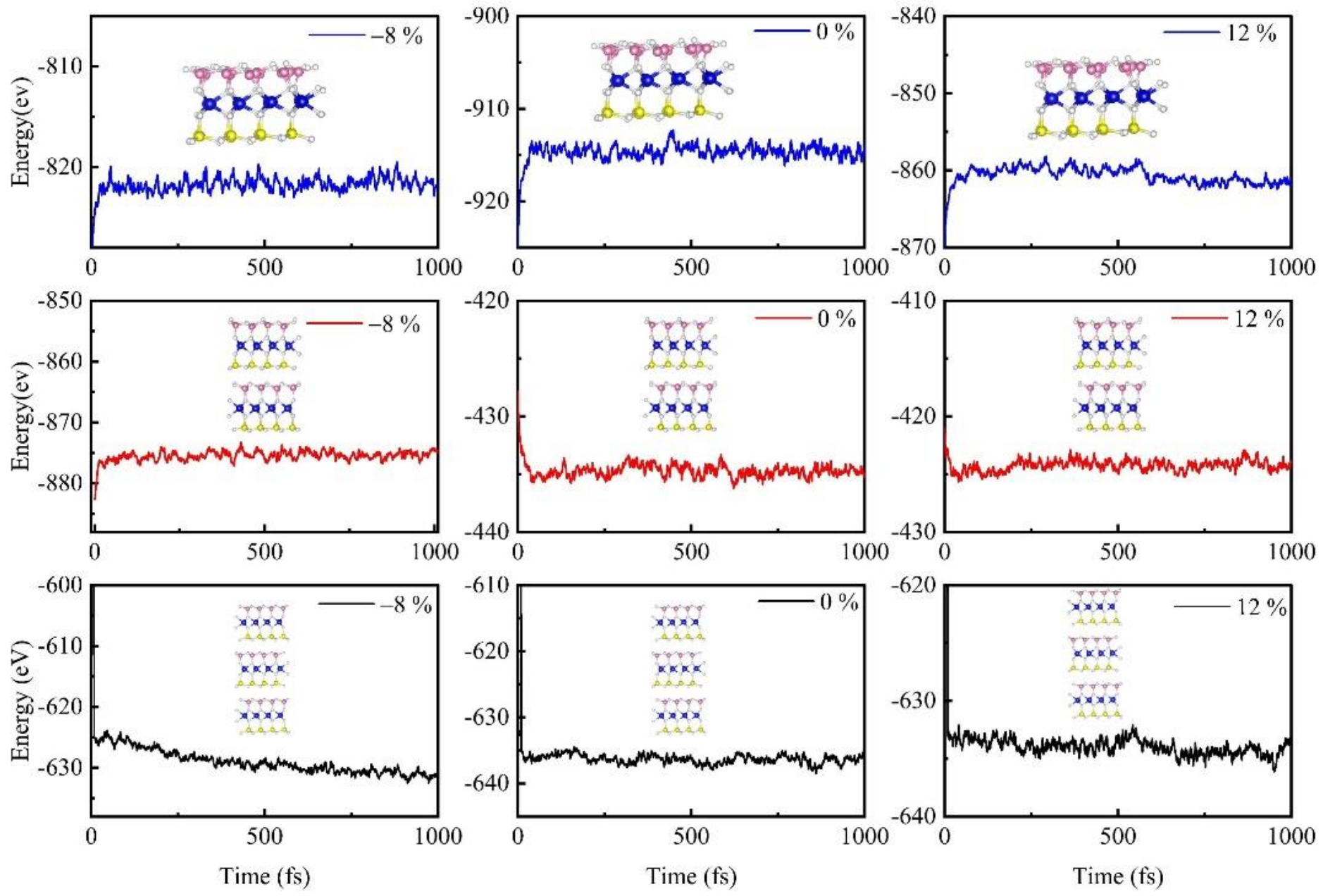

Fig. S4. The ab initio molecular dynamics (AIMD) of $MoSiGeN_4$ monolayer (blue line), bilayer (red line) and trilayer (black line) under different strains (−8%, 0% and 12%).

## (e) The orbital-projected band structure of $MoSiGeN_4$ monolayer

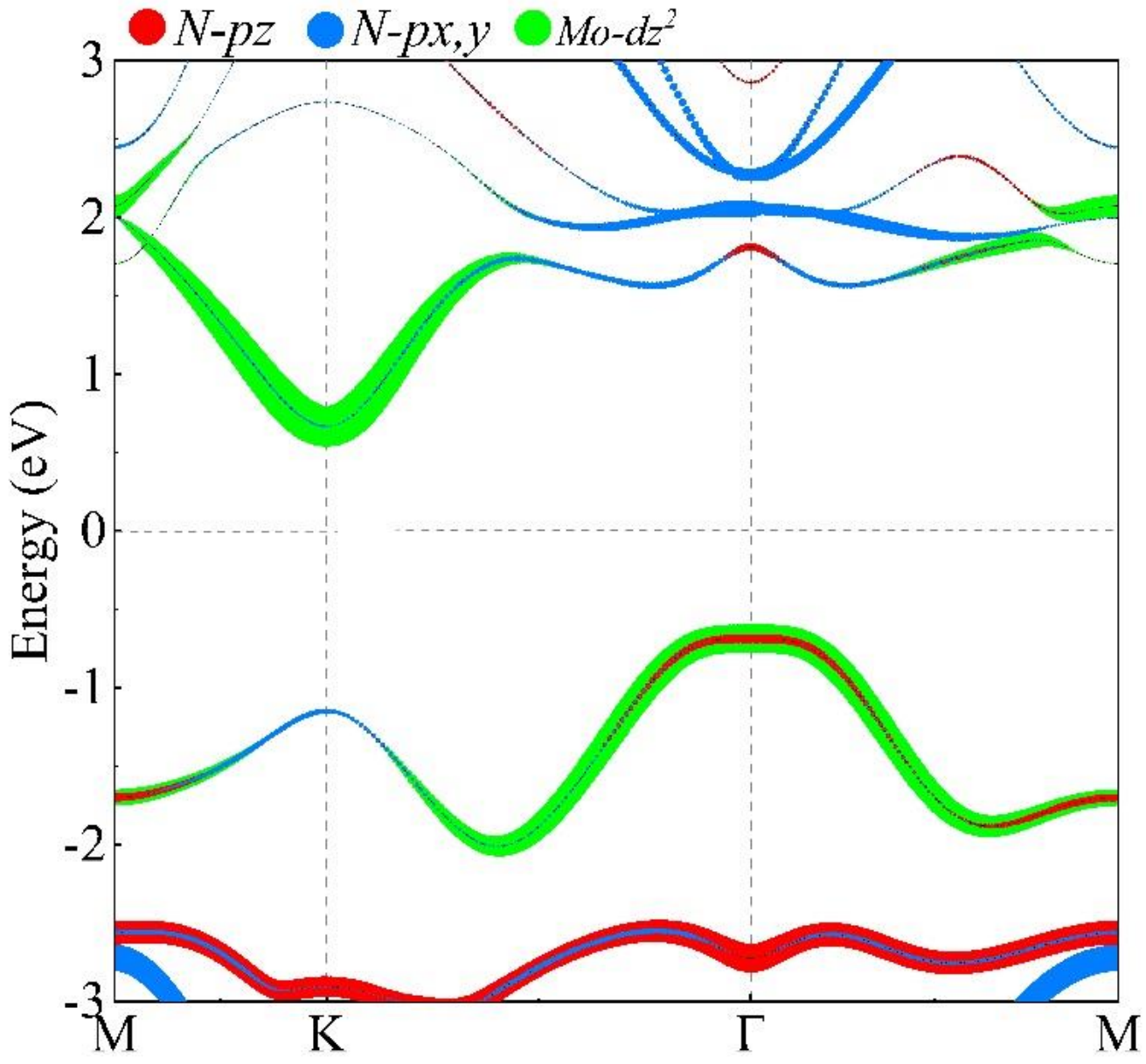

Fig. S5. The orbital-projected band structures of $MoSiGeN_4$ monolayer. The color of red is the contribution of N-$p_z$ orbital. The color of blue is the contribution of N-$p_{x,y}$ orbital. The color of green is the contribution of Mo-$dz^2$ orbital.

## (f) The orbital-projected band structures of $MoSiGeN_4$ mutilayer

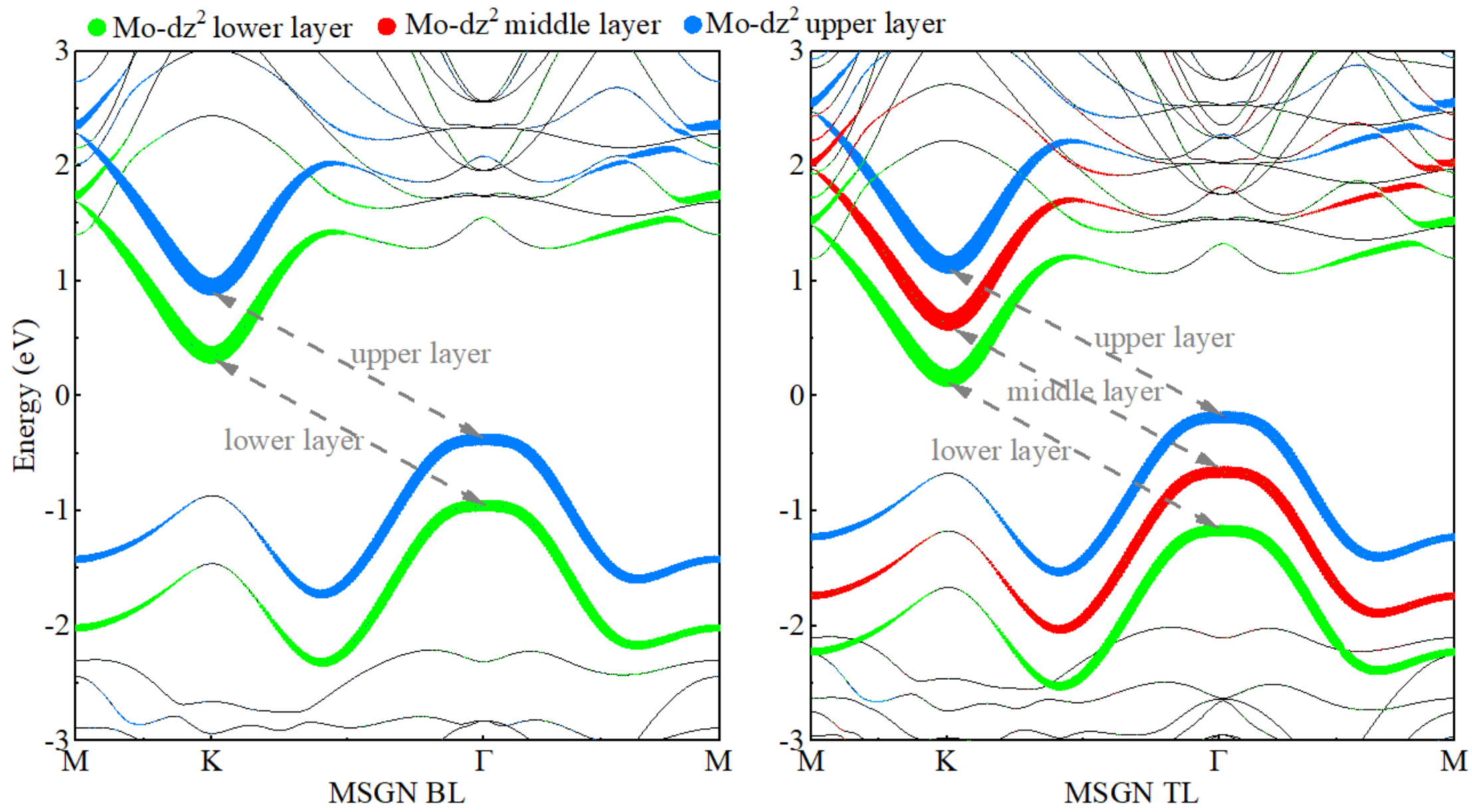

Fig. S6. The orbital-projected band structures of $MoSiGeN_4$ bilayer (MSGN BL) and $MoSiGeN_4$ trilayer (MSGN TL). The color of green is the contribution of Mo-$dz^2$ in the upper layer. The color of red is the contribution of Mo-$dz^2$ in the middle layer. The color of blue is the contribution of Mo-$dz^2$ in the lower layer.

## (g)The band structure of trilayer $MoSiGeN_4$

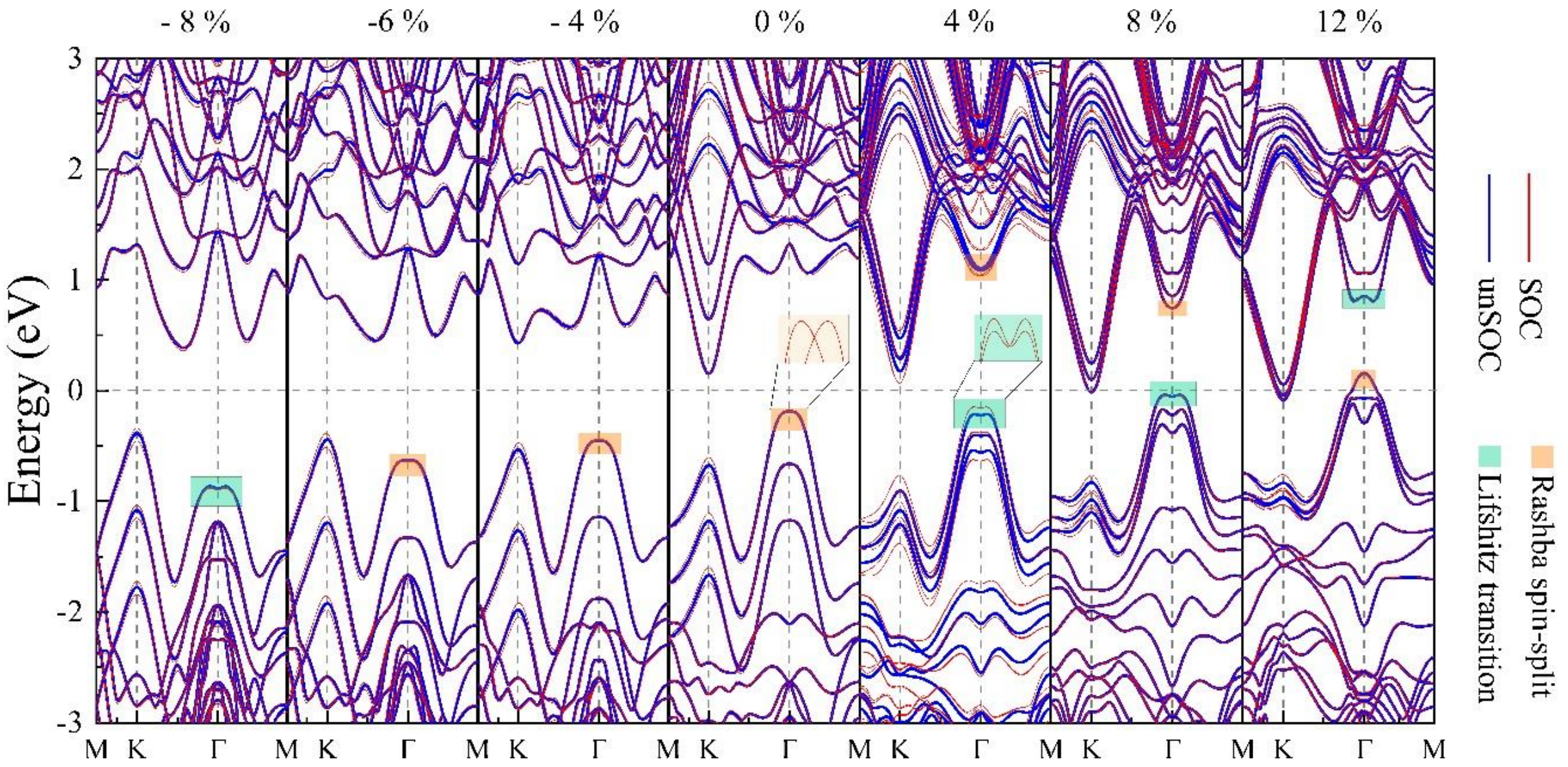

Fig. S7. Band structure of trilayer $MoSiGeN_4$ under biaxial strains considering without (blue lines) and with SOC (red lines). The enlarged drawing of Rashba spin splitting (orange patch) and Lifshitz transition (green patch) are shown in the inset.

## (h) The Rashba constants of $MoSiGeN_4$ monolayer and mutilayer

Table SI: The $k_0$, $E_R$, $\alpha_R$ and position of Rashba effect in $MoSiGeN_4$ monolayer, bilayer and trilayer under biaxial strains. $V_1$ denotes the highest valence band at Γ point, and $C_1$ denotes the lowest conduction band at Γ point.

| *Structure* | *Strain (%)* | *Position* ( Γ ) | $k_0$ (Å$^{-1}$) | $E_R$ (meV) | $\alpha_R$ (meVÅ) |
|---|---|---|---|---|---|
| *Monolayer* | -6 | $V_1$ | 0.110 | 5.317 | 96.401 |
| | -4 | $V_1$ | 0.063 | 1.204 | 37.897 |
| | -2 | $V_1$ | 0.049 | 0.764 | 30.690 |
| | 0 | $V_1$ | 0.070 | 2.151 | 61.073 |
| | 2 | $C_1$ | 0.078 | 5.220 | 134.290 |
| | 4 | $C_1$ | 0.027 | 1.831 | 135.179 |
| | 6 | $C_1$ | 0.033 | 2.332 | 140.355 |
| | 8 | $C_1$ | 0.067 | 6.638 | 195.898 |
| *Bilayer* | -6 | $V_1$ | 0.086 | 2.790 | 64.700 |
| | -4 | $V_1$ | 0.036 | 0.514 | 28.087 |
| | -2 | $V_1$ | 0.031 | 0.390 | 24.983 |
| | 0 | $V_1$ | 0.052 | 0.988 | 37.424 |
| | 2 | $C_1$ | 0.079 | 4.248 | 106.680 |
| | 4 | $C_1$ | 0.019 | 1.270 | 130.256 |
| | 6 | $C_1$ | 0.019 | 1.490 | 155.208 |
| | 8 | $C_1$ | 0.032 | 2.120 | 130.061 |
| *Trilayer* | -6 | $V_1$ | 0.075 | 2.370 | 62.434 |
| | -4 | $V_1$ | 0.042 | 0.405 | 19.148 |
| | -2 | $V_1$ | 0.024 | 0.250 | 20.815 |
| | 0 | $V_1$ | 0.062 | 0.627 | 19.968 |
| | 2 | $C_1$ | 0.059 | 3.552 | 118.459 |
| | 4 | $C_1$ | 0.017 | 1.307 | 154.583 |
| | 6 | $C_1$ | 0.019 | 1.563 | 163.152 |
| | 8 | $C_1$ | 0.093 | 15.108 | 324.449 |

## (i) Multilayer $MoSiGeN_4$ with applying biaxial strain without vdW

When the density functional dispersion correction (DFT-D3) is not considered, the electronic structure of monolayer is unchanged (as shown in Fig. S7). For bilayer and trilayer $MoSiGeN_4$, the interlayer spacing and electronic structure is different from considering DFT-D3. The interlayer distance in both bilayer and trilayer $MoSiGeN_4$ decrease drastically when biaxial strain change from 8% to 10%, meanwhile, a new band structure, $L_m$, appeared at 10%. In detail, when $\varepsilon \leq -8\%$, the Lifshitz transition is found. When -8% < $\varepsilon \leq 0\%$, Rashba spin splitting is revealed in the highest valence band of Γ point. When 0% < $\varepsilon \leq 9\%$, it evolves into the Lifshitz transition again, and when $\varepsilon = 10\%$, it has five lifshitz transition points. When $\varepsilon \geq 12\%$, Rashba spin splitting shows up again, exhibiting a L←R→L→$L_m$→R evolution near Fermi surface.

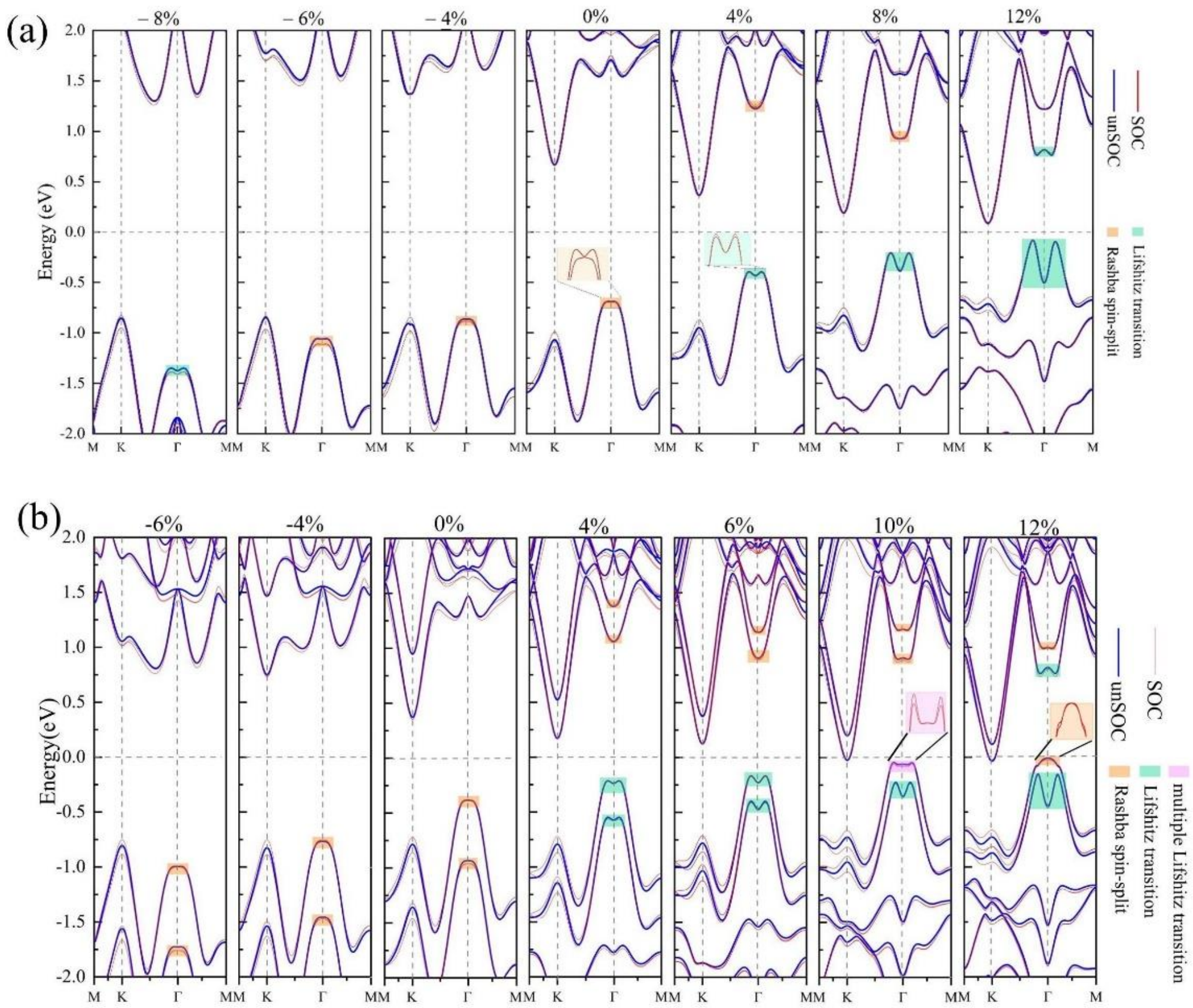

Fig. S8. Band structures of (a) monolayer and (b) bilayer $MoSiGeN_4$ under biaxial strains by using PBE (blue lines) and PBE+SOC (red lines) without vdW. The enlarged drawing of Rashba spin splitting (orange patch) and Lifshitz transition (green patch) are shown in the inset.

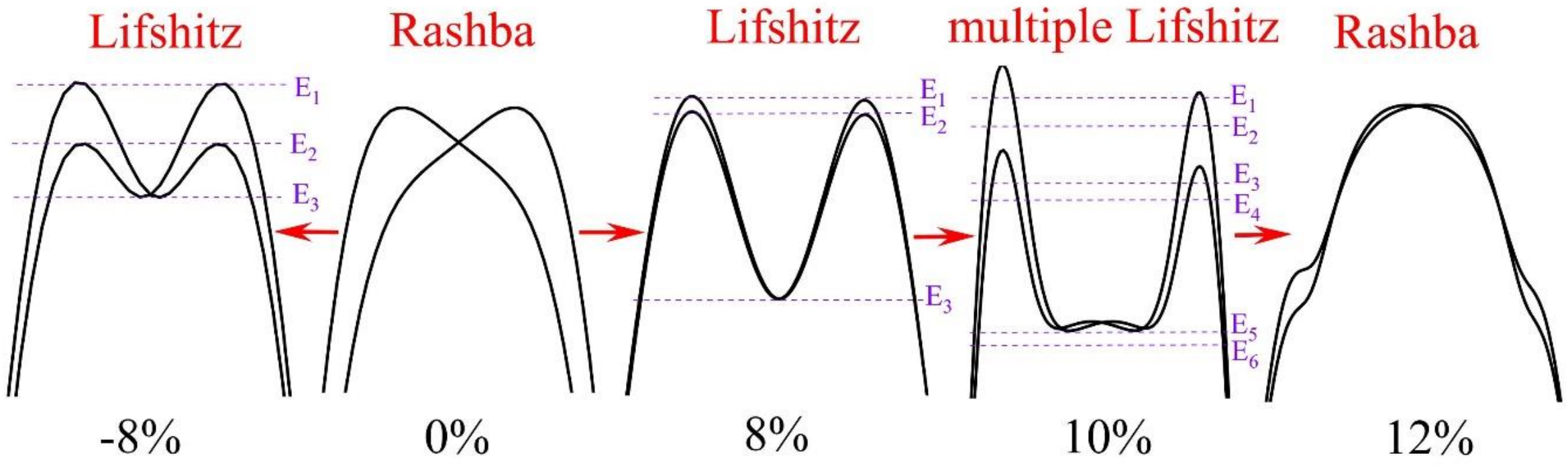

Fig. S9. The L←R→L→$L_m$→R evolution near the Fermi level of bilayer $MoSiGeN_4$ when DFT-D3 is not considered .